\pdfoutput=1
\documentclass[usenatbib, useAMS]{mnras}

\usepackage{graphicx}	% Including figure files
\usepackage{amsmath}	% Advanced maths commands
\usepackage{amssymb}	% Extra maths symbols
\usepackage{multicol}        % Multi-column entries in tables
\usepackage{bm}		% Bold maths symbols, including upright Greek
\usepackage{pdflscape}	% Landscape pages
\usepackage{wasysym}    % Astronomy Symbols
\usepackage{upgreek}    % Upright Greek Symbols
\usepackage{gensymb}    % General Symbols
\usepackage{orcidlink}  % orcid
\usepackage[T1]{fontenc}
\usepackage{ae,aecompl}

\newcommand{\orcid}[1]{\href{https://orcid.orffi#1}{\textcolor[HTML]{A6CE39}{\aiOrcid}}}

\begin{document}
% Title of the paper, and the short title which is used in the headers.
% Keep the title short and informative.
\title{Gas dynamical friction as a binary formation mechanism in AGN discs}

% The list of authors, and the short list which is used in the headers.
% If you need two or more lines of authors, add an extra line using \newauthor
\author[S. DeLaurentiis et al.]{
Stanislav DeLaurentiis$^{1}$\orcidlink{0000-0002-8922-825X}
\thanks{Contact e-mail: \href{mailto:sod2112@columbia.edu}{sod2112@columbia.edu}},
Marguerite Epstein-Martin$^{1}$\orcidlink{0000-0001-9310-7808}
\thanks{Contact e-mail: \href{mailto:mae2153@columbia.edu}{mae2153@columbia.edu}},
Zoltán Haiman$^{1,2}$\orcidlink{0000-0003-3633-5403}
\thanks{Contact e-mail: \href{mailto:zh2007@columbia.edu}{zh2007@columbia.edu}},
\\
% List of institutions
$^{1}$Department of Astronomy, Columbia University, 550 W. 120th Street, New York, NY
10027, USA\\
$^{2}$Department of Physics, Columbia University, 550 W. 120th Street, New York, NY
10027, USA}

% These dates will be filled out by the publisher
\date{}

% Enter the current year, for the copyright statements etc.
\pubyear{2023}

% Don't change these lines
\label{firstpage}
\pagerange{\pageref{firstpage}--\pageref{lastpage}}
\maketitle

\begin{abstract}
In this paper, we study how gaseous dynamical friction (DF) affects the motion of fly-by stellar-mass black holes (sBHs) embedded in active galactic nucleus (AGN) discs. We perform 3-body integrations of the interaction of two co-planar sBHs in nearby, initially circular orbits around the supermassive black hole (SMBH). We find that DF can facilitate the formation of gravitationally bound near-Keplerian binaries in AGN discs, and we delineate the discrete ranges of impact parameters and AGN disc parameters for which such captures occur. We also report trends in the bound binaries' eccentricity and sense of rotation (prograde or retrograde with respect to the background AGN disc) as a function of the impact parameter of the initial encounter.  While based on an approximate description of gaseous friction, our results suggest that binary formation in AGN discs should be common and may produce both prograde and retrograde, as well as both circular and eccentric binaries.
\end{abstract}


%%%%%%%%%%%%%%%%%%%%%%%%%%%%%%%%%%%%%%%%%%%%%%%%%%

%%%%%%%%%%%%%%%%% BODY OF PAPER %%%%%%%%%%%%%%%%%%

\section{Introduction}

Mergers between stellar-mass black holes (sBHs) detected by the LIGO/Virgo collaboration have revealed the existence of a population of binary black-holes (BBHs) \citep{2019Abbott, 2021abbott, 2021LIGO}. While their ultimate coalescence is observable via gravitational waves (GWs), the pathway by which these binaries form remains controversial. A variety of formation scenarios have been proposed, including, for example, isolated binary star evolution \citep{2016belczynski}, dynamical evolution of triple or quadruple systems \citep{2017silsbee, 2017liu, 2019liu, 2019fragione}, and chance encounters in dense stellar systems such as globular clusters and nuclear star clusters \citep{2009oleary, 2012antonini, 2017banerjee, 2018rodriguez, 2019fernandez}. 

In recent years, the possibility of AGN accretion discs as promising sites for BBH mergers has gained significant attention. Whether captured from the nuclear population \citep{2017bartos, 2018panamarev, 2020macleod, 2020fabj} or formed in situ \citep{2003levin, 2012mckernan, 2016stone}, AGN are hosts to the densest population of compact objects and stars in the universe \citep{2010merritt}. These embedded objects are expected to exchange torques with the gaseous disc and migrate inward, forming binaries via accumulation and close encounters in migration traps \citep{2016bellovary, secunda_orbital_2019, 2020secunda, Yang_2019}, in annular gaps~\citep{tagawa_formation_2020} or elsewhere in the disc through low-velocity interactions \citep{2012mckernan}.

Among the open questions is the origin of the AGN disc-embedded binaries in the first place. While the majority of massive stars in the Galaxy are in binaries, it is not clear whether the same is true for stars in AGN discs, forming under very different conditions~\citep{Cantiello+2021}.  On the other hand, single sBHs are expected to have many close interactions, due to their differential radial migration towards the central SMBH (with massive sBHs overtaking less massive ones). \citet{tagawa_formation_2020} suggested that during these close encounters the background AGN disc can provide friction, extracting enough relative kinetic energy to lead to capture. The process is analogous to the proposed formation of Kuiper-belt binaries by dynamical friction (DF) on the background ``pebbles" in our Solar system~\citep{goldreich_formation_2002}. Using simplified toy-models for the gaseous friction, combined with one-dimensional N-body simulations, \citet{tagawa_formation_2020} found that such ``gas-capture binaries" typically contribute the majority of all binaries in the AGN disc (i.e. out-numbering pre-existing binaries in the galactic nucleus that are captured by the disc). 

In this paper,  we further examine these proposed gas-capture binary formation events, using explicit orbital calculations which include the gaseous version of DF~\citep{ostriker_dynamical_1999}.  Our goal is to assess the conditions that lead to capture, and to examine the properties of the captured binaries: their initial separation, eccentricity, and sense of rotation.

Recent studies have addressed various aspects of this problem. \citet{boekholt_jacobi_2022} performed direct orbital integrations of 3-body systems, consisting of two small masses around a central massive body, without any dissipation.  They examined the orbits in detail, and mapped out the impact parameters leading to extremely close interactions and presumed binary capture through GW emission.  They found that the set of these impact parameters has a fractal structure, overall making capture exceedingly rare.

\citet{li_long-term_2022} also performed orbital integrations of multiple satellites around a SMBH, and incorporated friction forces through a toy model. They parameterised friction through a damping timescale (assumed to be much longer than the orbital time). They also found that binaries form in rare close encounters through gravitational wave emission, but the resulting weak friction did not enhance the rate of binary formation. They did not examine the orbital paths or the eccentricities of binaries post their formation.

During the completion of this manuscript, we became aware of closely related work by \citet{LiDempseyLi+2022}, who performed two-dimensional hydrodynamical simulations of binary formation in an AGN disc.   They examined only a single initial impact parameter, but a range of initial azimuthal separations and AGN disc densities.  They found successful binary formation in several cases, all of which produced eccentric and retrograde binaries.  They described a simple criterion for gas capture based on these runs. We compare our approach and results to their work in more detail in \autoref{ssec:compare}.

In this study, we follow the orbital evolution of AGN-disc embedded sBHs, systematically examine the captured binary orbits, and the dependence of capture on system parameters.  While our treatment is simplified, relying on an approximate treatment of gaseous dynamical friction, it is computationally relatively inexpensive, and allows us to study the properties of the orbits over a large parameter space.  We find successful captures that produce both eccentric and circular, and both prograde and retrograde binaries.  We also argue that capture is typically caused by friction on the approach towards (rather than during) close encounters, lending some confidence to our approximate treatment.

This paper is organised as follows.
In \autoref{sec:methods}, we discuss our simulation setup, AGN disc model, and computational infrastructure. 
In \autoref{sec:succ_caps}, we present examples of successfully captured orbits, along with their morphology, and discuss the role of dynamical friction in producing these captures.
In \autoref{sec:ecc_rot} we further investigate characteristics of captured binaries, focusing on the eccentricity and sense of rotation of their orbits,
and also determine the impact parameter ranges leading to capture.
In \autoref{sec:astrophysical_reality}, we discuss the dependence of our results on the background AGN disc density and the strength of the dynamical friction, as well as how these results can be extrapolated to parameter values beyond the simulated ranges. 

In \autoref{sec:implications}, we discuss the implications of our results for sBH binary formation in physical AGN disc models.
Finally, in \autoref{sec:conclusions} we summarise our conclusions and the implications of this work.

\section{Methods}\label{sec:methods}

\begin{figure}
    \centering
    \includegraphics[width=1\columnwidth]{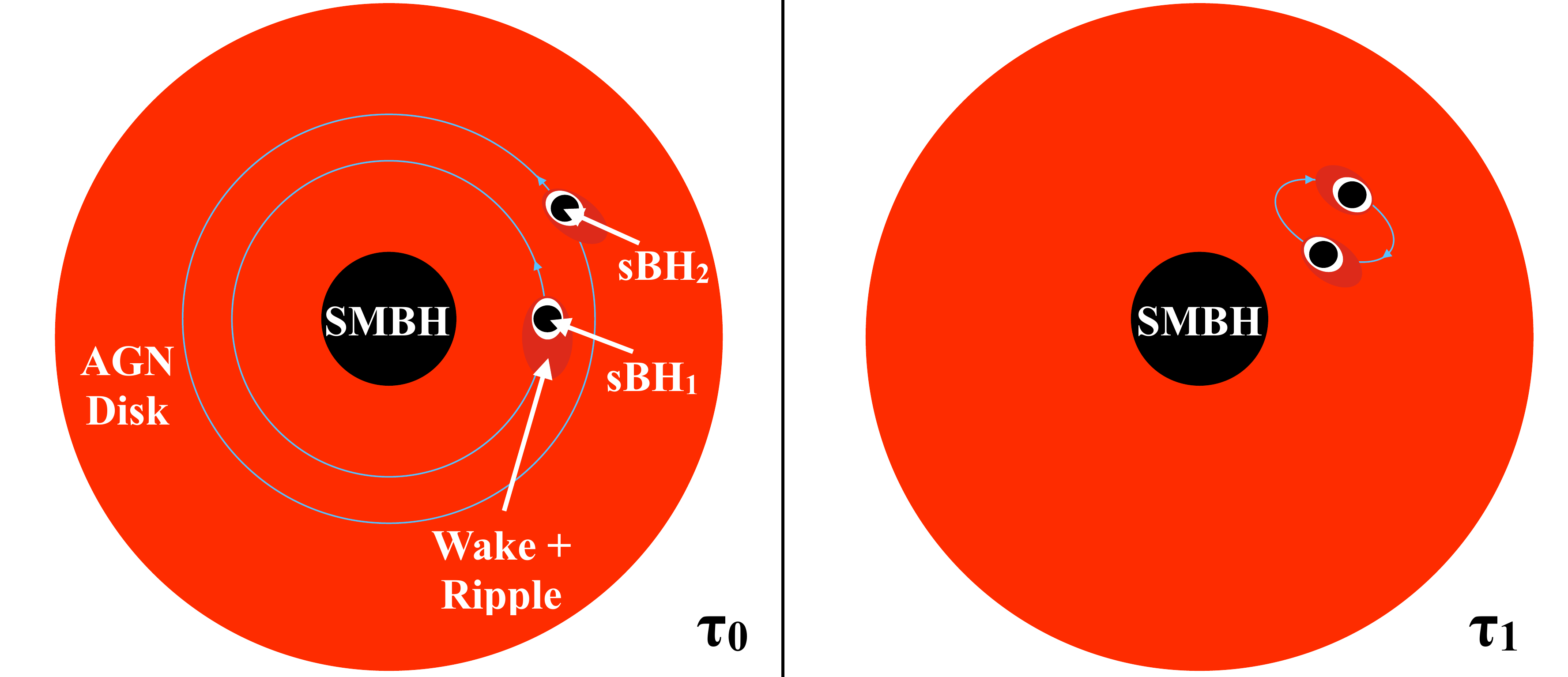}
    \caption{A schematic diagram showing our simulation setup. Our simulation is initialized (at $\tau_0$) with two sBHs in the disc, widely separated in azimuth and narrowly separated in radius, orbiting the SMBH at their respective Keplerian velocities. The sBHs' orbits are subsequently impacted by gas dynamical friction, and as they pass each other, they begin to dynamically interact, potentially forming a bound binary (at $\tau_1$) as a result of this friction.}
    \label{fig:df_diagram}
\end{figure}

This study models the close approach, co-planar motion of sBHs in an AGN disc by employing both 3-body gravitational dynamics and the \citet{ostriker_dynamical_1999} formulation of gaseous dynamical friction. The two-dimensional motion of each body is simulated in the lab frame with the adaptive n-body integrator \verb_REBOUND_ \verb_IAS15_, using the precision parameter \verb_Epsilon_ $=10^{-8}$ \citep{rein_rebound_2012,rein_2015_ias15}\footnote{We confirmed the numerical precision of this integrator by using it to successfully reproduce select orbits from \citet{goldreich_formation_2002} and \citet{boekholt_jacobi_2022}.}.

We begin simulating the system when the radially and azimuthally separated single sBHs are travelling around the SMBH at their Keplerian velocity, approaching each other on separate, initially circular, orbits. We integrate the equations of motion on Columbia's High Performance Computing Cluster \textit{Habanero}, accounting for gravity and gas DF at each time step (see \autoref{fig:df_diagram} for a schematic picture). We then stop our simulations when a bound binary is formed (see \autoref{ssec:experimental_design}) or when the simulation time exceeds twice the orbital period of the outermost sBH around the SMBH.

\subsection{AGN disc model}\label{ssec:discmethods}

\begin{table*}
\begin{tabular}{|c|c|c|}
 \hline
 Symbol& Description&  Fiducial Value \\
 \hline
 $M_{0}$   &  Mass of central SMBH & $10^{8} {\rm M}_{\astrosun}$\\
 $m_{1}$   &  Mass of inner sBH ($\rm{sBH}_{1}$) & $24 {\rm M}_{\astrosun}$\\
 $m_{2}$   &  Mass of outer sBH ($\rm{sBH}_{2}$) & $24 {\rm M}_{\astrosun}$\\
 $r_{0,1}$ & Initial distance from SMBH to $\textrm{sBH}_1$ & 0.1 pc\\
 $r_{0,2}$ & Initial distance from SMBH to $\textrm{sBH}_2$ & 0.1pc $+$ $b\sqrt[3]{3}R_{\rm Hill}$\\
 $b$ & Dimensionless impact parameter & [0,2]\\
 $\rho$   &  AGN disc gas density  & $10^{-12.3}$ ${\rm g\,cm^{-3}}=10^{9.9}{\rm M}_{\astrosun}\,{\rm pc^{-3}}$\\
 T   &  AGN disc gas temperature & $10^{4.5}$ [K]\\
 $\Delta \phi$ & Initial azimuthal separation between $\textrm{sBH}_1$ and $\textrm{sBH}_2$ & 10 $R_{\textrm{Hill}}$\\
 \hline
\end{tabular}
\caption{Model parameters and their fiducial values.}
\label{tab:fiducial_params}
\end{table*}

We assume the AGN disc to be a geometrically thin, optically thick, radiatively efficient, and steady-state disc \citep{Yang_2019}. We model this disc according to the \citet{sirko_spectral_2003} prescription\footnote{We find that the exact choice in AGN model is not critical to achieving capture (see \autoref{ssec:degenerate_orbits}).}, a modified \citet{shakura_sunyaev_disk_1973} disc with a constant accretion rate fixed at an Eddington ratio of $\dot{M}=L_{\rm Edd}/\epsilon c^2=0.5$, where $L_{\rm Edd}$ is the Eddington luminosity and $\epsilon=0.1$ is the assumed radiative efficiency.
Further, we assume that our gas is ideal and circling the SMBH at a constant Keplerian velocity. We set the adiabatic index to be that which is expected for an ideal approximation of the gas around a BBH, $\Gamma=5/3$ \citep{chapon_hydrodynamics_2013}. We also set the mean molecular weight $\upmu=1.15 m_{\rm p}$ appropriate to an ionized H+He gas, where $m_{\rm p}$ is the proton mass. 

\subsection{Dynamical friction}\label{ssec:dyfric}
A point-like object traveling in a collisionless, uniform medium with a constant velocity is decelerated by dynamical friction \citep{Chandra_og}. In our simulations, we adopt the equation for the gaseous version of dynamical friction as formulated by \citet{ostriker_dynamical_1999},
\begin{equation} \label{eqn:ostriker}
F_{DF}=\frac{-4\pi\\G^2M^2\rho}{v_{M}^3}f(\frac{v_{M}}{c_{s}})\boldsymbol{v_{M}}
\end{equation}
\begin{equation}\label{eqn:f(mach)}
f(x) = \left\{
        \begin{array}{ll}
            0.5\ln(\frac{1+x}{1-x})-x & 0 < x < 1 \\
            0.5\ln(x^2-1)+\ln(\lambda_{\rm C}) & x >1.
        \end{array}
    \right.
\end{equation}
Here $M$ is the mass of the body moving though the gas, $v_{M}$ is its velocity (relative to the Keplerian background gas of the AGN disc), $c_s$ is the sound speed and $\rho$ is the density of the gas, $\ln(\lambda_{\textrm{C}})=3.1$ is the Coulomb factor, and the argument of the function $f(x)$ is the (relative) orbital Mach number ${x\equiv v_M/c_s}$. The value of $\ln(\lambda_{\textrm{C}})=3.1$ was chosen to represent a BH moving through a hydrodynamic disc \citep{chapon_hydrodynamics_2013}.

\begin{figure}
    \centering
    \includegraphics[width=1\columnwidth]{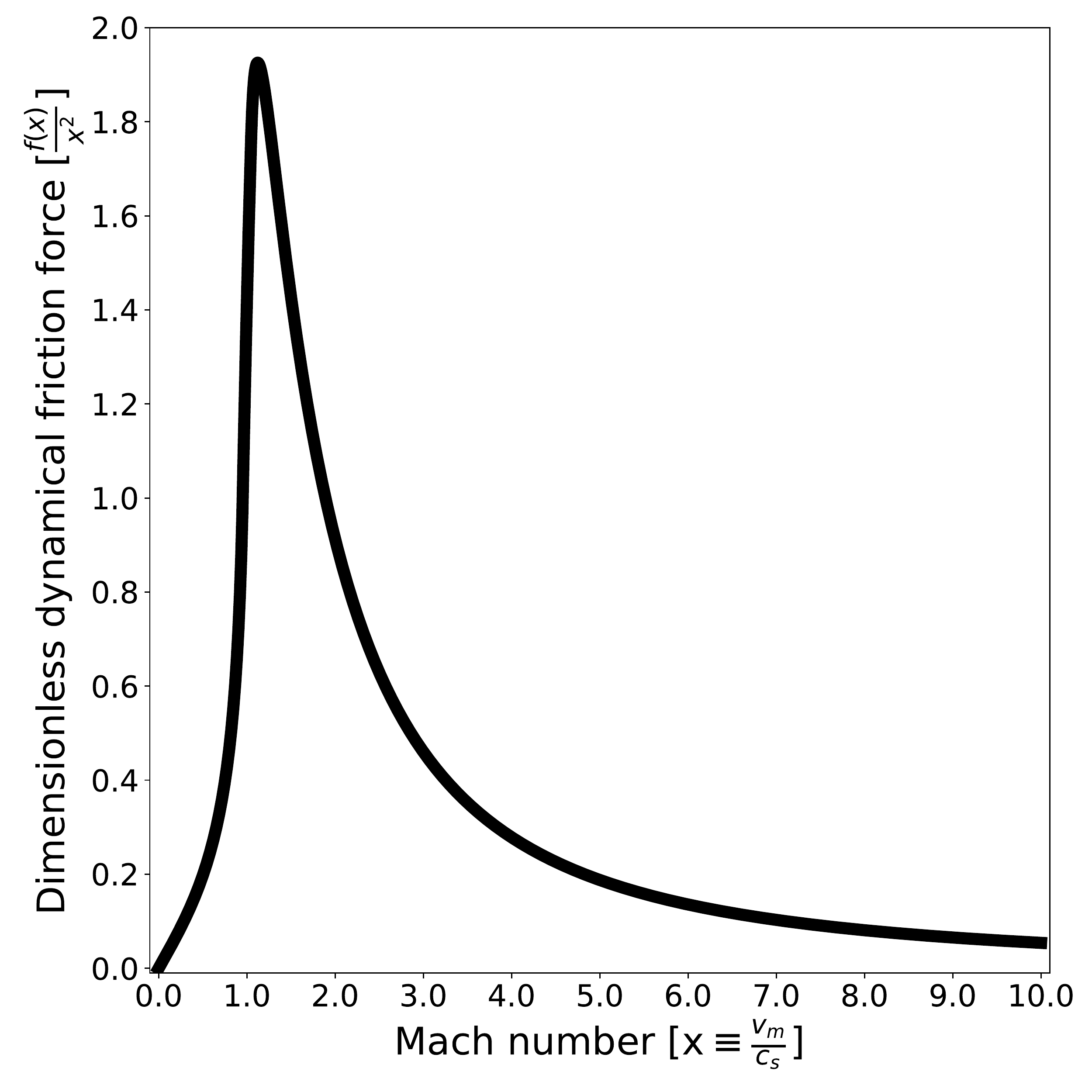}
    \caption{Dimensionless gas dynamical friction force versus Mach number used in our simulations (see \autoref{eqn:f(mach)}, adapted from \protect\citealt{ostriker_dynamical_1999}).  The force peaks at the slightly supersonic speed of $x=1.22$.}
    \label{fig:ostriker_machplot}
\end{figure}

Near $x=0$, \autoref{eqn:ostriker} is unbound. This is unphysical, since if a body is co-moving exactly with the gas (${x=\frac{v_M}{c_s}=0}$) there should be no wake, and no friction acting on the body. To account for this, we set $f(x)=0$ for small Mach numbers $x<10^{-4}$. Similarly, we also avoid the unphysical, unbound behavior at $f(x=1)$. Namely, we apply a linear approximation from $x=1-10^{-2}$ to $x=1+10^{-2}$ to create a continuous function around $f(x=1)$. The resulting prescription of DF used in our simulations is shown as a function of Mach number in \autoref{fig:ostriker_machplot}. 

To confirm the validity of our use of a uniform-density medium, we note that the radial length-scale along which the AGN disk density varies, $\rho/(d\rho/dr)=r/(d\ln\rho/d\ln r)$ is order $\sim r$ for power-law density profiles $\rho \propto r^{\alpha}$.  This is much larger than both the Hill radius and the Bondi radius.
On the other hand, our prescription for dynamical friction assumes a disc with scale height $H>R_{\rm{Hill}}$ and Bondi radius $R_{\rm{Bondi}} < R_{\rm{Hill}}$. While these conditions hold in our fiducial model, we discuss other parameter cases when these conditions are not satisfied in \autoref{ssec:capturediscs}.

\subsection{Experimental design}\label{ssec:experimental_design}
We investigate the role of DF in binary capture via two main experimental setups. 

First, we run a suite of simulations in a fiducial model (see \autoref{tab:fiducial_params}). The fiducial model uses sBH masses similar to the progenitors of LIGO/Virgo BBH mergers \citep{2021abbott} and places them in the outer region of a typical AGN disc where stellar-mass compact objects are expected to be abundant \citep{tagawa_formation_2020}. The background gas parameters (density and temperature) listed in \autoref{tab:fiducial_params} are directly adopted from the \citet{sirko_spectral_2003} AGN disc profile with a $10^{8} {\rm M}_{\astrosun}$ central SMBH, as shown in Figure~1 in \citet{secunda_orbital_2019}.

In this study, the fiducial model serves two purposes. First, it is a reference model, which we can compare to variants with different parameter choices, in order to understand the mechanics of binary capture. Second, within the fiducial model, we can study (a) the capture occurrence as a function of the impact parameter $b$, as well as basic properties of the successfully captured binaries, such as their (b) initial orbital separations, (c) eccentricities, and (d) sense of rotation.

More specifically, to understand the dependence of our results on $b$, we systematically vary its value in the range $0 < b < 2$, running simulations at intervals of 0.01, followed by smaller intervals of $10^{-3}$ and $10^{-4}$ to investigate particular smaller ranges of interest, as discussed below.

In our second setup, we again simulate interacting close pairs of sBH orbits, as depicted in \autoref{fig:df_diagram}, but now we vary parameters such as BH mass and disc temperature and density, in addition to $b$, in order to understand the parameter combinations that result in capture (see \autoref{sec:astrophysical_reality}).

For both setups, we define ``capture" as the formation of a bound sBH binary. In practice, after some experimentation, we adopted the following specific capture criteria in our fiducial model: the semi-major axis of the orbit of the two sBHs around one another (in the co-rotating frame) is less than $\Delta b=0.2$ and has changed by less than $5\%$ for 10 consecutive sBH binary orbits. In a second experimental setup, we contrast this with a simpler capture criterion that is easier to implement and does not require as long a simulation run, namely: when the total (potential + kinetic) energy of the binary, treated as an isolated system, is below a certain threshold (see \autoref{ssec:cap_tol}).

\subsection{Initial conditions and binary orbital parameters}

In this paper, we define the Hill radius ($R_{\textrm{Hill}}$) based on the initial position of the outer sBH ($\textrm{sBH}_2$),
\begin{equation}
\label{eqn:r_hill}
R_{\textrm{Hill}}=r_{0,2}\sqrt[3]{{\frac{m_{2}}{3M_{0}}}}.
\end{equation}

The initial azimuthal separation of the two sBHs is ${\Delta \phi = 10 R_{\textrm{Hill}}}$, measured along the arc of the orbit of sBH$_2$.
The initial radial separation between the two sBH orbits, which we also refer to as the impact parameter, is defined in terms of the dimensionless coefficient, $b$
\begin{equation}
\label{eqn:dr_hill}
r_{0,2}-r_{0,1}=b(\sqrt[3]{3})R_{\textrm{Hill}}
\end{equation}

If a run results in capture, we proceed to compute the captured binary's orbital parameters. Namely, we measure both the semi-major and semi-minor axis of the orbit, by identifying the first pericenter and apocenter, in the co-rotating center of mass frame (see \autoref{ssec:morphology}) after satisfying the capture criteria.   We also measure the time evolution of the sBH binary orbit's azimuthal coordinate ($\phi$) to determine the sense of rotation (i.e. prograde or retrograde with respect to the background AGN disc's motion).

\section{Successful captures}\label{sec:succ_caps}

In this section we present the results of our fiducial model (see \autoref{tab:fiducial_params}). In \autoref{ssec:morphology} we demonstrate that under dynamical friction fly-bys can form BBHs. In \autoref{ssec:timetracing} we then show that the outcome of the encounter (i.e. capture or not) is determined before a close interaction occurs, i.e. before the sBHs begin orbiting around their mutual center of mass. In \autoref{ssec:density_vary} we study how the gas density (or more generally the strength of the dynamical friction) affects the range of impact parameters for which capture occurs, as well as some properties of the captured sBH binary orbits, such as their size, eccentricity, and precession. In \autoref{ssec:cap_regions} we determine the values of $b$ that result in capture.

\subsection{Orbital morphology}\label{ssec:morphology}

\begin{figure*}
    \centering
    \includegraphics[width=1\textwidth]{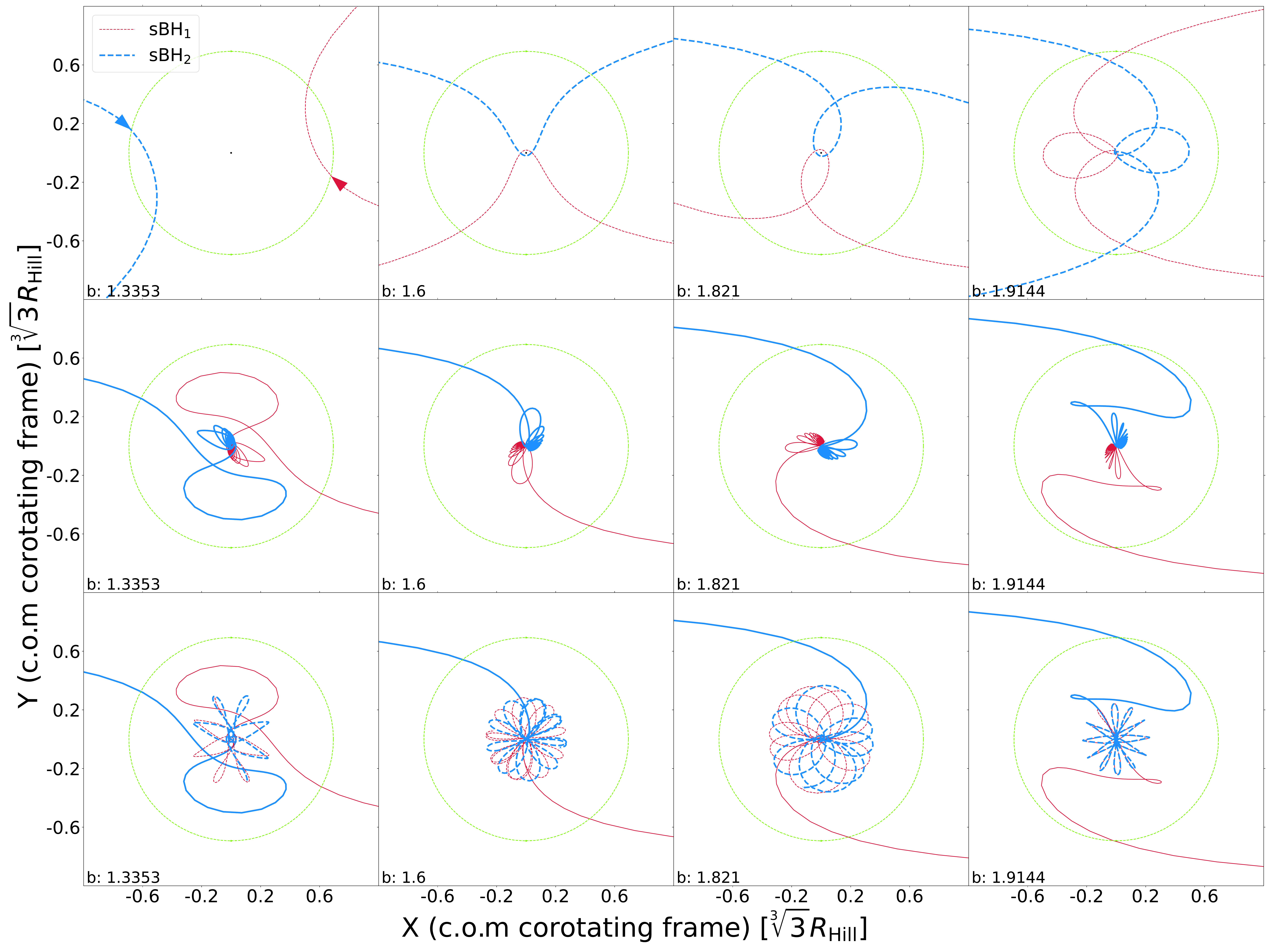}
    \caption{Select orbital paths, with dynamical friction off (top row), on (middle row), and on until just after the sBHs' first pericenter (bottom row). The solid lines depict where fiducial dynamical friction was on and the dashed lines depict where friction was off.The orbits are shown in the co-rotating frame, with the inner sBH$_1$ (red curve) entering from the lower right and the outer sBH$_2$ (blue curve) entering from the upper left [indicated by corresponding arrows]. The SMBH is located far below the plots at $(x,y)=(0, -0.1{\rm pc})$. The columns correspond to impact parameters (i.e. initial radial separations) of $b=1.3353, 1.6, 1.821, 1.9144$ (left to right; measured in units of Hill radii). Select impact parameters were chosen to illustrate the diversity of orbital shapes. The top row shows examples of fly-by's  without dynamical friction which instead result in binary captures when dynamical friction is turned on (bottom row). The distances are shown in units of $\sqrt[3]{3}R_{\textrm{Hill}}$ (see eq.~\ref{eqn:r_hill}) and the lime-green dashed circle marks the Hill radius around the center of mass.}
    \label{fig:fric_onvsoff_reg}
\end{figure*}

In \autoref{fig:fric_onvsoff_reg} we illustrate the orbital paths of the sBHs in a co-rotating frame.
More precisely, the frame is centered on the two sBHs' mutual center of mass, as they orbit about the central SMBH. The $y$ and $x$ axes of this frame then respectively correspond to the vectors along and perpendicular to a line connecting the SMBH and the sBHs' center of mass. The SMBH is located at $(x,y)=(0,-0.1{\rm pc})$.  The blue path corresponds to the outer sBH ($\rm{sBH}_{2}$) while the red path corresponds to that of the inner sBH ($\rm{sBH}_{1}$). Initially the sBHs follow the AGN disc gas along the Keplerian shear flow, so that in this frame $\rm{sBH}_{1}$ and $\rm{sBH}_{2}$ enter from the lower right and upper left, respectively. We define prograde and retrograde sBH binary motion as those with angular momenta parallel and anti-parallel to that of the background AGN disc, corresponding to clockwise and anti-clockwise motion in the co-rotating frame, respectively.

The orbital paths depicted in \autoref{fig:fric_onvsoff_reg} are of select simulations run with (middle row) and without (top row) dynamical friction. The columns are organized by impact parameter with values selected to illustrate the wide variety of our results. The middle panels in this figure show that DF can capture sBHs into bound binary orbits. Unlike the frictionless paths in the top row, where sBHs have one close encounter (``fly-bys"), DF brings sBHs together to sustainably orbit each other, yielding many close encounters. The orbits depicted in the middle row never untangle. Rather, they continue to shrink and become more bound due to the continuous loss of kinetic energy from DF. Eventually, GW emission (not included in this model) would cause the captured binary to coalesce. Our simulations with DF do not suggest the existence of ``temporary" binaries that orbit their barycenter but are eventually pulled apart by tidal forces (as can occur without DF; \citealt{boekholt_jacobi_2022}). Rather, under DF sBHs are either captured in bound binaries, or miss each other and do not orbit around their barycenter at all.

We also note the diversity of orbital shapes of the captured binaries that dynamical friction can produce. Orbits that undershoot and overshoot the center of mass (c.o.m.) can both result in capture. Despite differences in orbital paths, we observe that all captured orbits exhibit a slowing prograde precession as well as a slowing (with respect to binary orbital frequency) rate of semi-major axis shrinkage, resulting in orbits that, when traced over time, begin to stack on top of each other in \autoref{fig:fric_onvsoff_reg}.

\subsection{The role of dynamical friction for captures}\label{ssec:timetracing}

The aim of this work is to assess the effect of DF on binary sBH capture in an AGN disc. In an effort to better understand the role of DF in capture, we ran a suite of simulations (using fiducial parameters) in which DF was turned off following the first close approach of the sBHs. The resulting trajectories are illustrated in the bottom panels of \autoref{fig:fric_onvsoff_reg}, while the trajectories that had fiducial DF on throughout the run are depicted in the middle panels of \autoref{fig:fric_onvsoff_reg}.

\begin{figure}
    \centering
    \includegraphics[width=1\columnwidth]{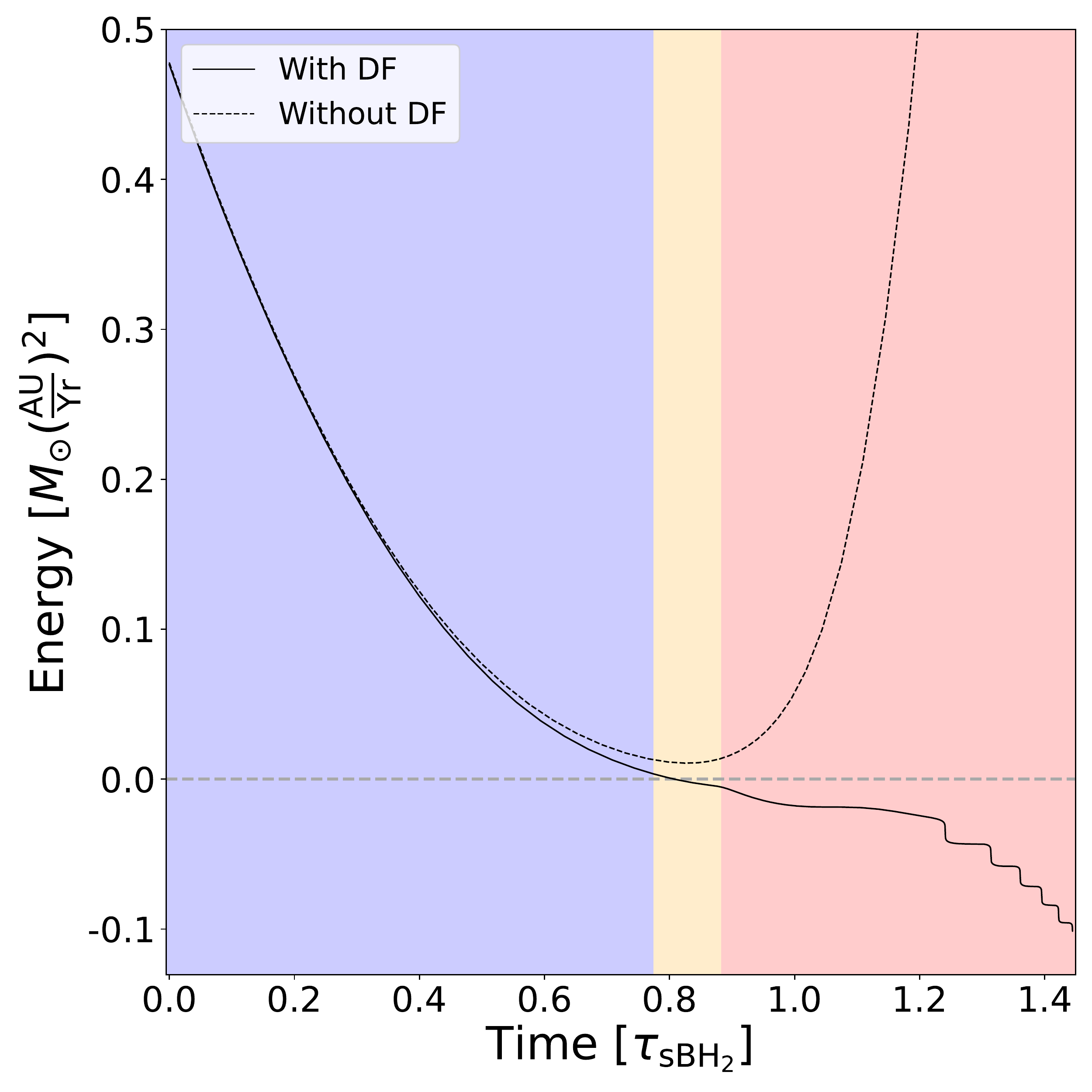}
    \caption{Evolution of the nominal binary orbital energy for a typical encounter between the two sBHs resulting in capture with impact parameter $b = 1.3361$ in our fiducial simulation suite. The solid line depicts the energy of the orbit when DF is turned on, and the dashed line depicts the energy of the orbit when DF is turned off, and the binary is no longer captured. Time is in units of the orbital period of $\rm{sBH}_2$ around the SMBH. The blue shading depicts the region of the orbit before the two sBHs enter their mutual Hill sphere, the orange shading depicts the orbit within the mutual Hill sphere but before the sBHs' first pericenter passage around their mutual center of mass, and the red shading depicts the remainder of the orbit.}
    \label{fig:capture_energyloss}
\end{figure}

The friction-less section of the orbits shown in the bottom panel of \autoref{fig:fric_onvsoff_reg} share several interesting characteristics. First and foremost, turning friction off does not disrupt the binary. Instead, the sBHs continue to orbit one another, maintaining a constant rate of orbital precession and a constant semi-major axis. The importance of friction during the initial sBH close approach is further illustrated in \autoref{fig:capture_energyloss}, in which we have plotted the binary orbital energy\footnote{We study the energy of the system in the two-body limit. This energy is only meaningful when the bodies have a small separation, and act like a two-body system.} with and without DF over time for an encounter with impact parameter $b = 1.3361$. Note that the DF and friction-less simulations begin with the same energy and both experience a drop in energy because of the change in the bodies' positions. They only significantly deviate from each other just before they enter the Hill sphere. Further, we note that the binary with DF becomes marginally bound $E=0$ when the body just enters the Hill sphere, and then falls well below zero prior to reaching pericenter. Subsequent binary orbits result in further energy loss, although less significant than the first encounter. This is unlike our friction-less case that never dips below $E=0$ during the entire orbit and does not yield a bound binary.

Further, we note that sBHs that are nominally only marginally bound by this naive criterion ($E \approx 0$) are typically not captured into bound orbits resembling isolated binaries.  Instead, they often ``fly-by'' each other, either directly, or in some cases after a single close encounter. This suggests that binaries do not become bound during their close interactions inside their mutual Hill sphere, but rather, sufficient energy loss from DF during the initial approach is what determines capture and binary formation.

Since binary capture is determined during the initial approach, this suggests that our DF prescription, though relatively simple, may be sufficient to establish and characterize DF-facilitated orbital capture. However, we note that our prescriptive use of \citet{ostriker_dynamical_1999} friction is only valid as long as the sBH orbital trajectories remain approximately linear, and is violated when the sBHs start strongly interacting and orbiting their mutual center of mass. At this point, we expect the binary wakes to interact and form circumbinary discs whereupon the binary's orbit can widen or decay, depending on the mass ratio, binary eccentricity, disc thickness, and sense of rotation.  Addressing the details of the orbital evolution is beyond the scope of the present work, as it requires hydrodynamical simulations. Such simulations have been performed both for isolated binaries~\citep{Miranda+2017,Tang+2017,Munoz_2019,Moody+2019,Tiede+2020,HeathNixon2020,Duffell+2020,DorazioDuffell2021,Franchini+2021} and recently also for stellar-mass BH binaries embedded in AGN disks~\citep{LiDempseyLi+2021,LiDempsey+2022,LiLai2022,LiLai2022b, Dempsey+2022}. We proceed with our analysis with the expectation that it will hold up in a hydrodynamic study, although this remains to be confirmed.

\subsection{Density variations}\label{ssec:density_vary}

\begin{figure*}
    \centering
    \includegraphics[width=1\textwidth]{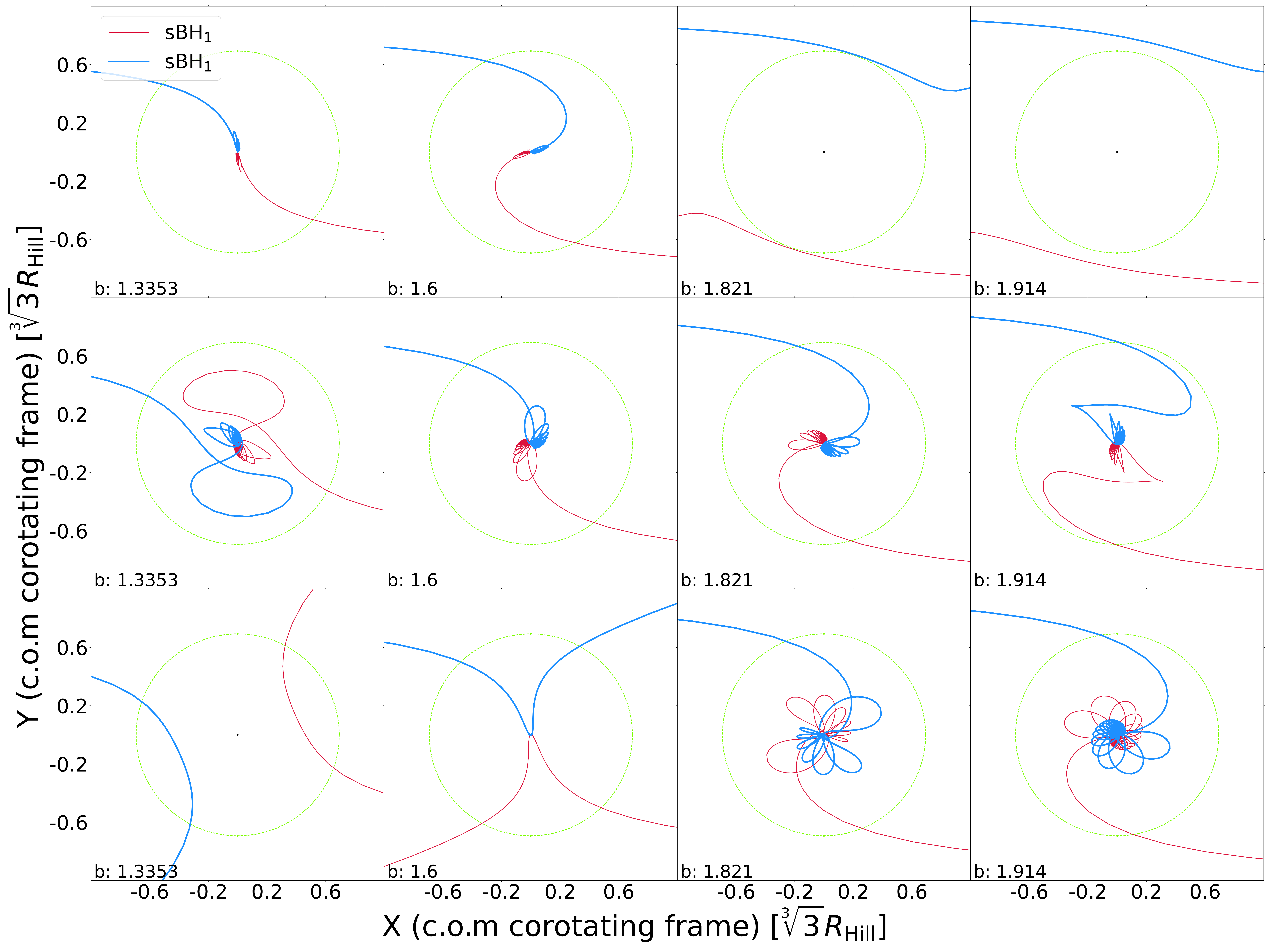}
    \caption{Similar to \autoref{fig:fric_onvsoff_reg}, but we show the orbital paths in simulations with varying background gas density. The middle row depicts runs with the fiducial density, $\rho=10^{-12.3} \rm{g\,cm^{-3}}$, and the top/bottom rows correspond to factors of 3 higher/lower densities of $10^{-11.8}$ and $10^{-12.8}\, \rm{g\,cm^{-3}}$, respectively.}
    \label{fig:fric_3up3down}
\end{figure*}

We carried out two suites of simulations that increased and decreased the fiducial disc density by a factor of 3, respectively, while keeping the SMBH mass, sBH masses, and disc temperature at their fiducial values. We note that the strength of dynamical friction simply scales linearly with $\rho$ (\autoref{eqn:ostriker}).

In addition to the parameters $\rho$, $\rm{v}_{\rm{M}}$, and $M$, DF is also a function of the Mach number (\autoref{eqn:f(mach)}). However, we note that our simulated sBHs generally have quite subsonic Mach numbers throughout their orbit ($x<0.5$). In combination with \autoref{fig:ostriker_machplot} this suggests that the force of dynamical friction itself only varies by at most 20\% due to these subsonic Mach number variations. This then indicates that Mach number variations generally do not significantly contribute to differences across various orbits, but rather the orbits of sBHs are ultimately governed by other parameters, such as $\rho$. Although there are exceptions, we confirm this understanding in \autoref{ssec:cap_tol} below. 

DF acts to keep bodies on Keplerian paths, tied to the assumed background gas. In the limit of extremely high density (high friction), the sBHs will be stuck, co-moving with the gas, and not interact. In the opposite limit of extremely low density, the sBHs will travel on nearly friction-less paths, and will not lose enough kinetic energy to form binaries. In this latter case, capture can still occur, but only for certain fine-tuned impact parameters resulting in ultra-close encounters and dissipation by gravitational waves (see \citealt{boekholt_jacobi_2022} for this so-called ``Jacobi capture"). We thus expect DF to greatly enhance the parameter space for binary capture, in those intermediate cases when the relative DF and mutual gravitational forces between the sBHs are comparable.

In \autoref{fig:fric_3up3down} we illustrate the competing effects of gravity and DF by comparing the orbital evolution of interacting sBHs embedded in gas of varying density. From top to the bottom panel, the densities are ${\rho = 10^{-11.8}}$, $10^{-12.3}$, $10^{-12.8} \rm{g\,cm^{-3}}$. Moving from left to right the density is constant, but the impact parameter is increased and thus the magnitude of mutual sBH gravity diminishes. In the top row, where density is at its highest, lower impact parameters (${b = 1.3353}$ and $1.6$) result in capture while higher impact parameters (${b = 1.821}$ and $1.914$) result in flybys. When $\rho$ is high, only at small impact parameters can sBH mutual gravity overcome DF and lead to capture. Note too that the semi-major axes of bound orbits in the high-density regime are significantly smaller than their lower density counterparts. The relative hardness of these binaries is a natural result of higher energy loss due to stronger DF.

In the bottom row of \autoref{fig:fric_3up3down} where $\rho$ is at its lowest, smaller impact parameters lead to flybys -- the inverse of the top row. Note that in the $b=1.6$ case, this is despite a very close interaction. This behavior can be explained by energy arguments, i.e. at low impact parameter and low $\rho$ there is not sufficient energy lost via dynamical friction for the orbit to become bound. At larger impact parameters, however, the path over which dynamical friction acts is long enough for the energy loss to exceed the binding energy and capture occurs. Clearly, $b$ and $\rho$ are two parameters that impact whether capture occurs, and both have ranges of values resulting in capture. These define a ``capture region'' in the $\rho,b$ plane. In the following section, we explore the ranges of $b$ producing captures in the fiducial simulation.

\subsection{Capture regions}\label{ssec:cap_regions}

In order to identify the impact parameters resulting in capture, we ran the fiducial simulations with iteratively sampled impact parameter values. We began with a ``sparse sample" of $0\leq b\leq 2$ with steps of $\Delta b=10^{-2}$.
After determining which impact parameters resulted in captured binaries, and whether these binaries experienced prograde or retrograde rotation, we ran two further sets of successively denser sampling ($\Delta b=10^{-3}$ and $10^{-4}$) around impact parameters where we saw large changes in capture or sense of rotation over a narrow range of $b$. These simulations were then used to determine which impact parameters result in capture, and to characterize the eccentricity and sense of rotation of the captured binaries (see \autoref{sec:ecc_rot}).

The impact parameters that captured sBHs into long-lived binaries in our fiducial model are found to cover three distinct bands. Namely at a resolution of up to $\Delta b=10^{-4}$,
any choice in $b$ within the closed intervals $b\in[1.335,1.3723]$, $[1.4672, 1.8986]$, and $[1.9036, 1.9148]$ saw sBHs captured. These bands
have uneven widths and are also unevenly spaced, as shown in magenta and green in \autoref{fig:rot_capture_hist} (this figure will be discussed in more detail in \autoref{ssec:eccrot_rh} below).

Our discussion of the impact of $\rho$ and $b$ as parameters directly correlated through DF and mutual gravity may suggest a single, continuous region of capture. However, the appearance of discrete ``capture bands" is not unique to this study. \citet{goldreich_formation_2002} performed approximate orbital calculations in the Hill approximation, reporting successful captures. On closer scrutiny, there is a close agreement between the capture bands visible in the upper left panel of their Figure 1 and ours, with both featuring three distinct bands---two narrow, one wide. \citet{boekholt_jacobi_2022}, who performed orbital calculations without DF, also observed ``islands" of capture, illustrated in their Figure 5. 

The capture bands in these papers are along similar impact parameter ranges. If we were to convert the \citet{boekholt_jacobi_2022} choice in impact parameters to ours, their regions of capture would be centered around $b=1.32$, 1.6, and 1.7. Though not exactly aligned with our bands we do see that they are in the same neighborhood (albeit much narrower and exhibiting fractal characteristics).

\section{Binary properties}\label{sec:ecc_rot}
In this section we present the results of our fiducial model, focusing on the captured orbits' eccentricity and sense of rotation. In \autoref{ssec:orbits_ecc} we demonstrate that dynamical friction can form binaries that are eccentric or circular, and rotating prograde or retrograde with respect to the background AGN gas. In \autoref{ssec:ecc_energy}, we demonstrate that DF dissipates energy differently in eccentric and circular orbits. In \autoref{ssec:eccrot_rh} we characterize fiducial binaries' eccentricity and sense of rotation as a function of the impact parameter.

\subsection{Eccentricity and sense of rotation}\label{ssec:orbits_ecc}

\begin{figure}
    \centering
    \includegraphics[width=1\columnwidth]{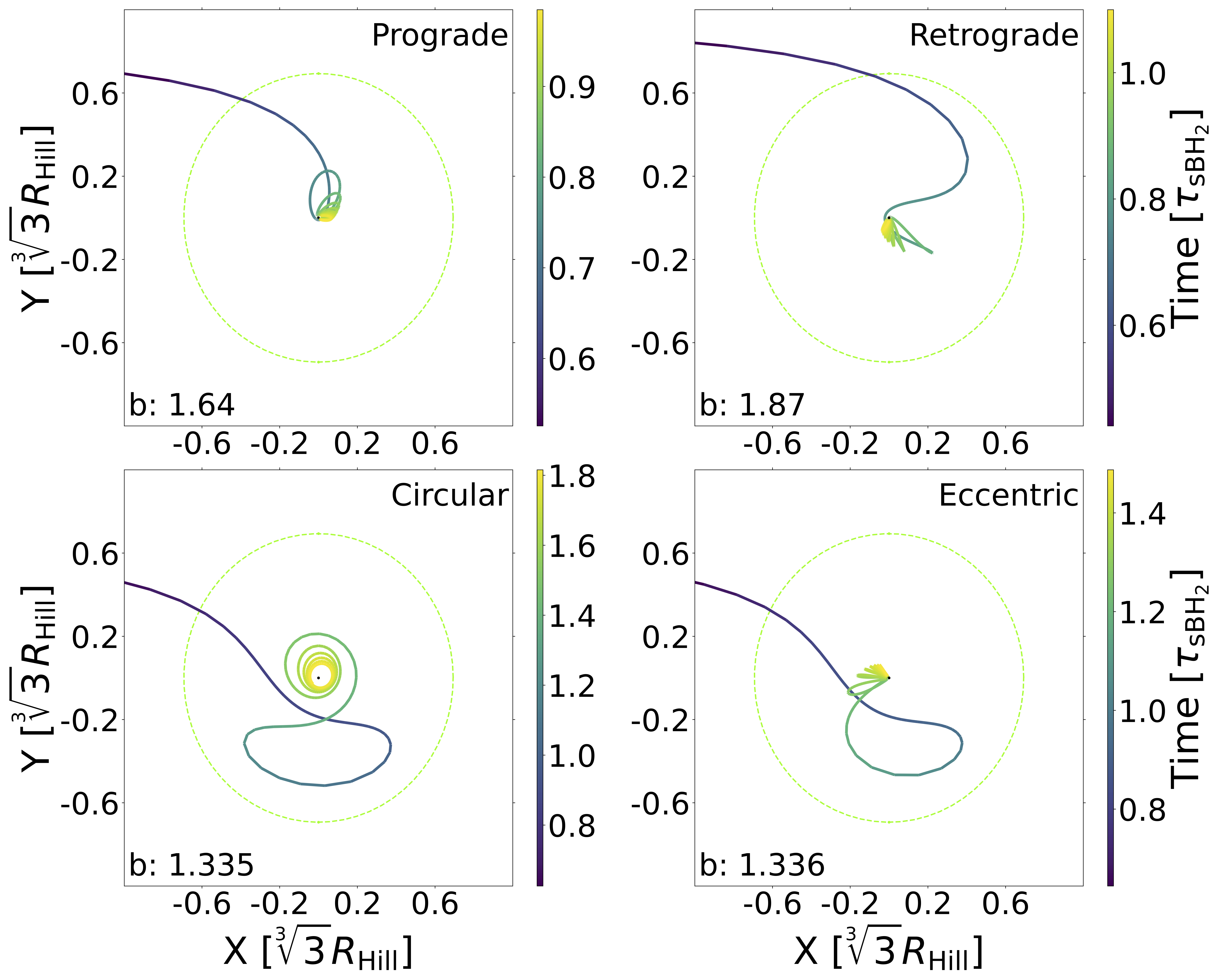}
    \caption{Orbital paths of select binaries, captured in the fiducial model. The top row contrasts examples of prograde {\it vs.} retrograde binaries ($b=1.64$, $1.87$) while the bottom row contrasts a circular {\it vs.} an eccentric binary ($b=1.335$, $1.336$).
    We note that the paths of $\rm{sBH}_2$ and $\rm{sBH}_1$ are mirrored in this center of mass co-rotating frame, thus (for clarity) we only show the path of $\rm{sBH}_2$. The orbital path is colored according to the orbital time of $\rm{sBH}_2$ around the SMBH. The frame is identical to that of \autoref{fig:fric_onvsoff_reg}}.
    \label{fig:shape_rot_panel}
\end{figure}

Eccentricity and sense of rotation are fundamental aspects of captured binaries. A few of the BH mergers discovered by LIGO/Virgo appear to prefer non-zero eccentricities~\citep{Romero-Shaw+2022}, and such residual eccentricity in the LIGO/Virgo band is a key signature distinguishing AGN disc-related from many other sBH binary formation pathways~\citep{tagawa_eccentric_2021, samsing_agn_2022}. Note that the eccentricities are caused by frequent binary-single interactions near the time of the binary's merger in these models. Without such interactions, eccentric sBH binaries formed due to gas dynamics are generally expected to 
be driven to an eccentricity of $e\sim 0.5$~\citep{Munoz_2019,Zrake_2021,DorazioDuffell2021}. However, after GWs dominate their inspiral, these binaries would circularize by the time they enter the LIGO/Virgo GW band. Non-negligible eccentricities can still exist in the LISA band, however \citep{Zrake_2021}. The sense of rotation of a captured binary, on the other hand, plays an important role during earlier stages, determining how binaries interact with circumbinary discs that are expected to form around them. In general, circumbinary discs are expected to be prograde, unless the binary's orbit around the SMBH itself is eccentric~\citep{LiChen+2022}.  Binaries that are co- {\it vs.} counter-rotating with respect to their circumbinary discs have different orbital evolutions and accretion rates -- for example, counter-rotating binaries are driven to merger much more rapidly~\citep{Nixon+2011, LiLai2022}.

In \autoref{fig:shape_rot_panel} we plot the orbital paths\footnote{While the simulations illustrated were run until the fiducial stopping conditions, we only depict parts of the orbit in this figure.} of four different binaries with four different impact parameters in our fiducial models. The figure shows that eccentricity is not constant, but rather evolves throughout the entire orbit, becoming either more circular or more eccentric with time ($b=1.335$ and $b=1.336$, respectively). We note that a binary's sense of rotation, however remains unchanged throughout its orbit. This is clearly illustrated in the upper rows of \autoref{fig:shape_rot_panel} ($b=1.64$ and $b=1.87$).

We further note that eccentricity and sense of rotation are not always correlated, as we find examples of both prograde and retrograde binaries that are eccentric (upper panels of \autoref{fig:shape_rot_panel}) as well as both prograde and retrogade binaries that are circular (the latter shown in the bottom left panel of \autoref{fig:shape_rot_panel}). In the following sections, we analyze the expected eccentricity evolution based on the energy dissipated by DF along the orbit, and examine eccentricity and sense of rotation as coupled functions of the impact parameter $b$.

\subsection{Eccentricity as energy loss}\label{ssec:ecc_energy}
\begin{figure}
    \centering
    \includegraphics[width=1\columnwidth]{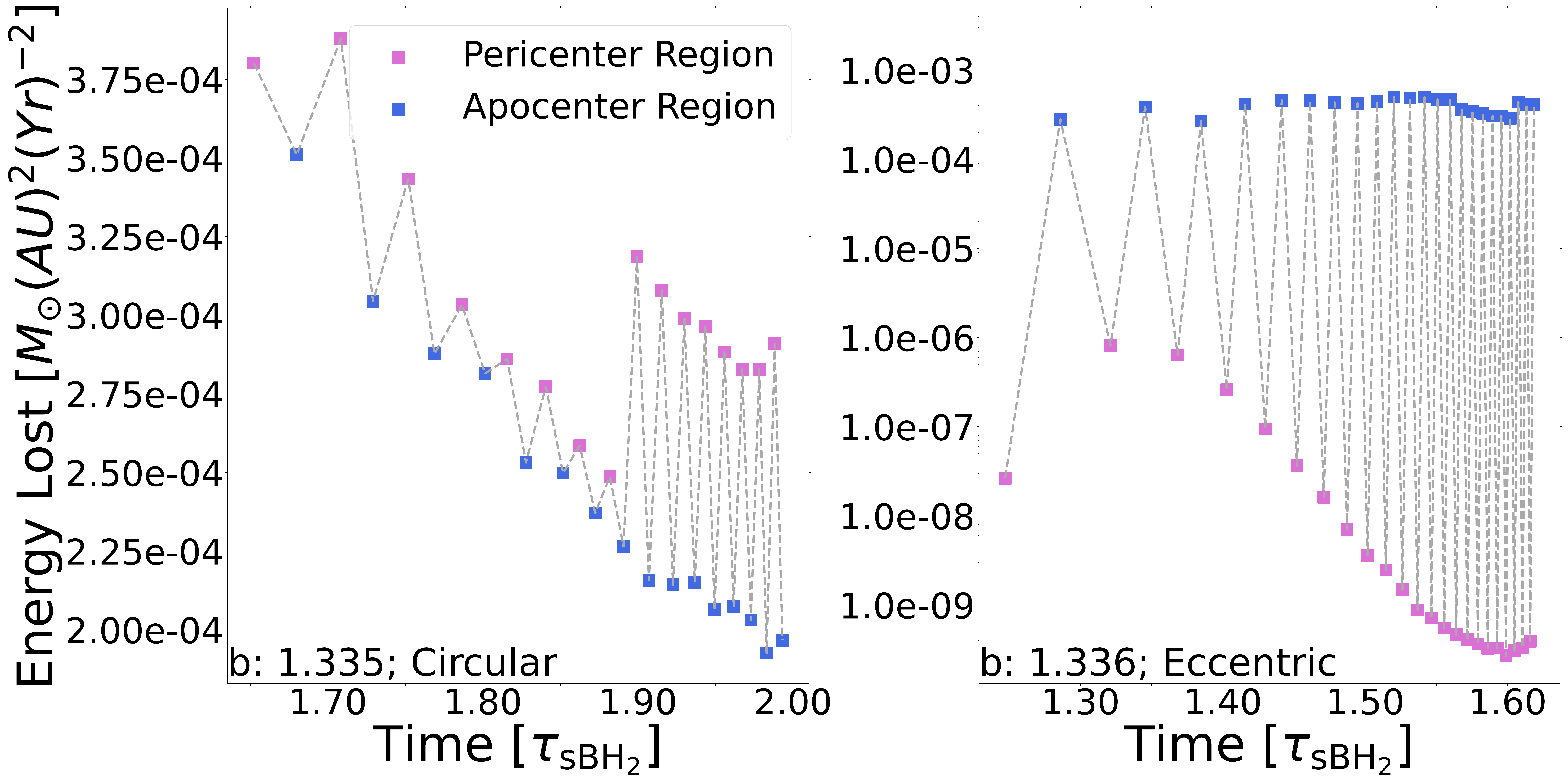}
    \caption{Energy of $\rm{sBH}_{2}$ dissipated by DF during its orbit inside the Hill sphere. Select fiducial suite simulations that show circular and eccentric binaries (left and right, $b=1.335$ and $b=1.336$) are depicted. The $x$ axis shows time in units of the orbital period of $\rm{sBH}_{2}$ around the SMBH ($\tau_{\rm{sBH}_{2}}$), and the $y$ axis shows the work done by DF. The purple markers depict the energy lost near the binary's pericenter and the blue markers depict the work done by DF near the binary's apocenter. The time value of the markers is the time of each pericenter or apocenter.}
    \label{fig:energyloss}
\end{figure}

\autoref{fig:energyloss} traces the energy dissipated\footnote{Energy lost was computed as the work done by DF during the $\pi/8$ sector before the binary's pericenter and apocenter, respectively. Taking the $\pi/8$ sector before or after the pericenter and apocenter did not significantly impact our results.} by DF near apocenter and pericenter as a function of time for select eccentric and circular binaries ($b=1.335$ and $b=1.336$), whose respective orbital paths are traced in the lower panels of \autoref{fig:shape_rot_panel}.

The eccentricity (e) of a bound binary is uniquely determined by its energy (E) and angular momentum (L) as
\begin{equation}
    {e}^2 = 1 + \frac{2E(m_1+m_2)L^{2}}{G^{2}(m_1 m_2)^{3}}.
\end{equation}
Note that $E<0$ but $L>0$. The dissipation of energy (i.e. $E$ becoming more negative, or larger $|E|$) in the system leads to a more circular binary, while a decrease in angular momentum (smaller $L$) leads to a more eccentric binary. In the impulse approximation of a non-conservative force, an impulse applied at the apocenter imparts a larger torque and change to the angular momentum than the same impulse applied at the pericenter. In the binary that becomes circular (\autoref{fig:shape_rot_panel} left panel) DF does more work near the pericenter, dissipating significant energy and minimal angular momentum, causing the binary to become more circular. In the eccentric orbit (\autoref{fig:shape_rot_panel} right panel) DF does more work near the apocenter, decreasing the angular momentum significantly and making the binary more eccentric. In our simulations, we find that greater energy loss near the apocenter or pericenter regions is always associated with eccentric or circular binaries, respectively. Further, we note that energy loss, generally, remains uneven throughout the entirety of our simulated orbits, driving binaries to form either nearly perfect circles or extremely high eccentricity ellipses at later times (as depicted by the bottom panels of \autoref{fig:shape_rot_panel}). Finally, we observe that binary eccentricity evolution is not dependent on the absolute value of the work done by DF, but rather only the relative energy lost through different phases of the orbit.

\subsection{Eccentricity and rotation as functions of impact parameter}\label{ssec:eccrot_rh}

\begin{figure*}
    \centering
    \includegraphics[width=1\textwidth]{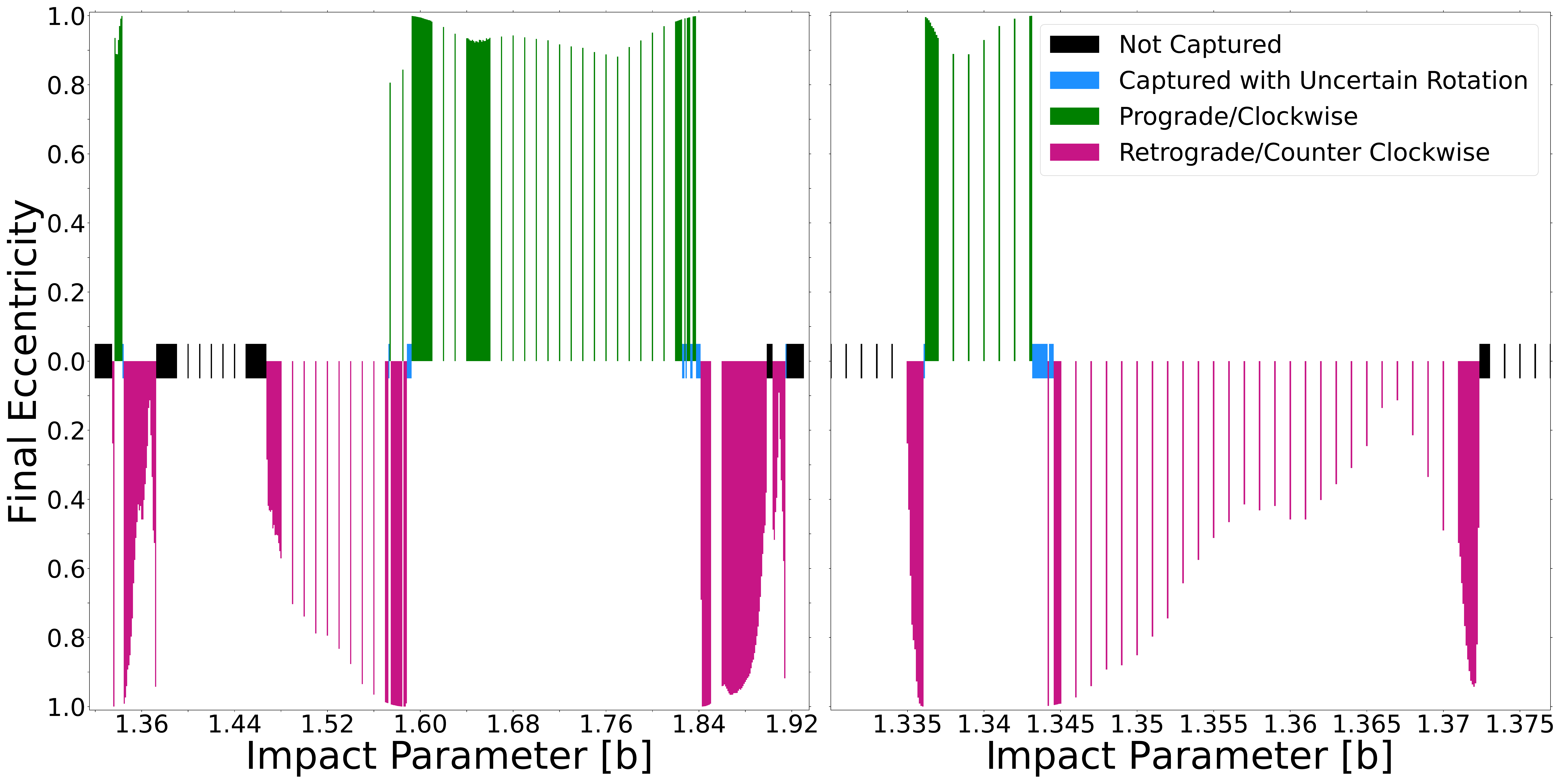}.
    \caption{The sense of rotation and eccentricities
    of orbits from our fiducial suite of simulations. The left panel depicts all captures, while the right panel is a zoomed in picture of the first ``capture band''. The $x$ axis represents the impact parameter and the $y$ axis represents the eccentricity of the last complete orbit before the simulation stopping condition (see \autoref{ssec:experimental_design}). Retrograde orbits are plotted in magenta along the negative $y$-axis while prograde orbits in green along the positive $y$-axis. Captured orbits with undetermined features are depicted in blue, orbits that are not captured are shown in black, and white space indicates unsampled regions in $b$. The width of each bar is $\Delta b = 10^{-3}$ (left panel) and $\Delta b = 10^{-4}$ (right panel) where $b=\sqrt[3]{3}R_{\rm{Hill}}$.}
    \label{fig:rot_capture_hist}
\end{figure*}

In \autoref{fig:rot_capture_hist} we show the eccentricity and rotation of captured orbits in our fiducial suite of simulations. The results of each simulation are plotted as a vertical line, whose height corresponds to the binary's orbital eccentricity just before the simulation was stopped. Prograde orbits are shown in green and plotted along the positive y-axis to differentiate them from their retro-grade counterparts, which are plotted in red along the negative y-axis.

In our suite of fiducial simulations, retrograde orbits make up $63\%$ of all captured orbits. This falls just within the range of eccentricity distributions found for long-lived binary interactions in the friction-less case, suggesting a similar equipartition between prograde and retrograde orbits in the case with DF \citep{boekholt_jacobi_2022}. 

We find that ``highly-eccentric'' orbits ($e\ge 0.8$) make up $74\%$ of all captured orbits, whereas ``near-circular'' ($e\le 0.2$) orbits only make up only $0.7\%$. We do not have an equipartition between circular and eccentric binary orbits. Moreover, the eccentricity distribution of binaries is asymmetric with respect to the binary's rotation. We find that all prograde orbits are ``highly eccentric'', while retrograde orbits occupy a wide range of eccentricities $e\in [0.1,1-10^{-5}]$.

There are several trends in eccentricity and rotation with respect to $b$, which are clearly illustrated in  \autoref{fig:rot_capture_hist}. We find that these trends hold true in all sampled parameter regimes (see \autoref{ssec:cap_tol}). 

\begin{figure}
    \centering
    \includegraphics[width=1\columnwidth]{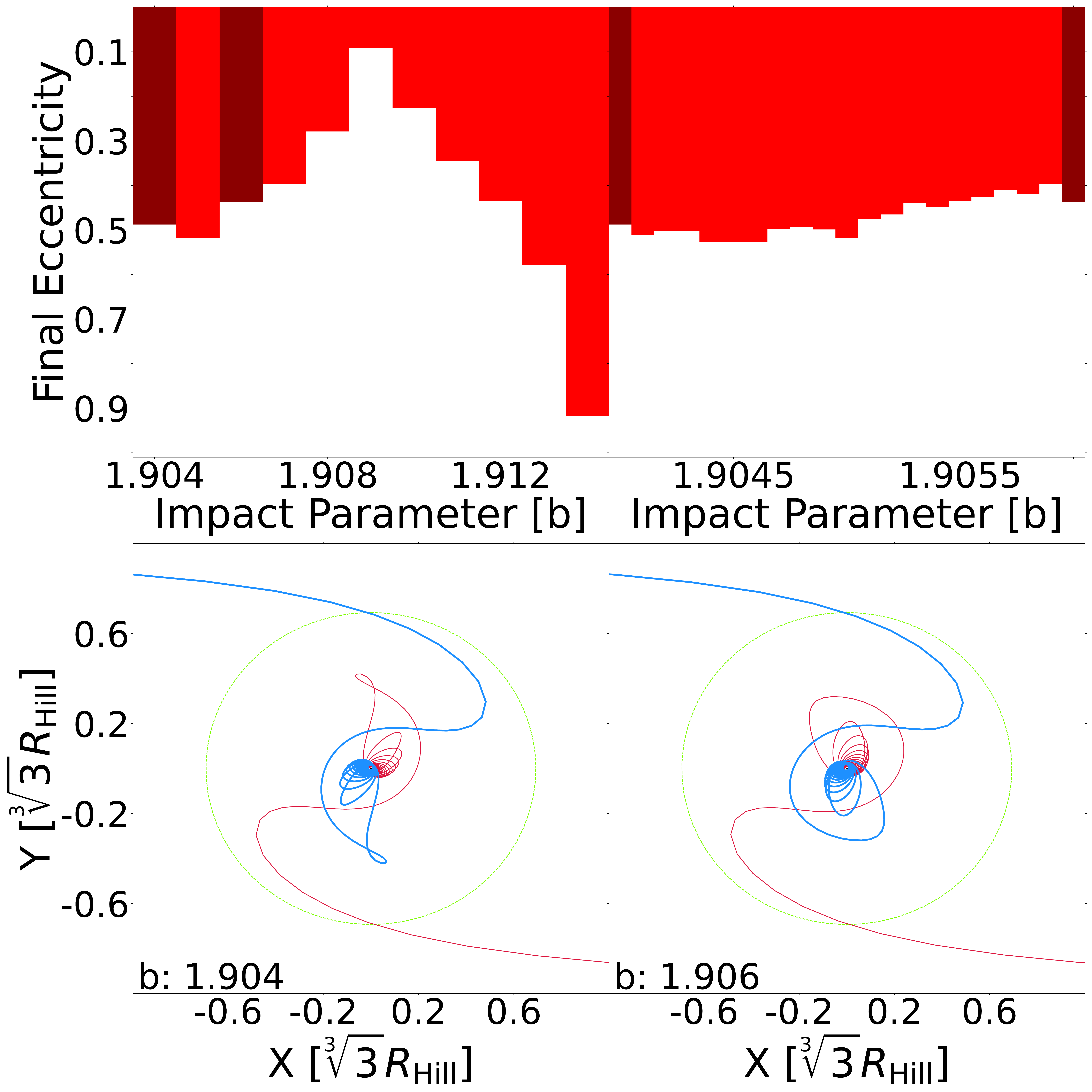}
    \caption{A zoom in around the simulations of the $b\in [1.904,1.914]$ region. In the top row we show the eccentricities of these select retrograde orbits, at increasing resolution (left $10^{-3}b$, right $10^{-4}b$). In the bottom row, we plot the orbital paths of two select simulations ($b=1.904,1.906$, brown columns in top row) in the center of mass co-rotating frame to illustrate their similarity.}
    \label{fig:no_fractal}
\end{figure}

\begin{enumerate}
\item{
Just within boundaries of capture, we observe low eccentricity orbits. This trend is clearly seen in \autoref{fig:rot_capture_hist}, which provides a close up of the capture band $b\in [1.335,1.3723]$. Though bordered by high eccentricity binaries in the interior of the band, the boundary cases have characteristically low eccentricities, $e=0.23,0.48$ for $b=1.335, 1.3723$. That is, marginally captured objects tend to form circular binaries. Though these circular orbits may not be resolved for all capture bands in \autoref{fig:rot_capture_hist}, we have confirmed that under sufficiently high resolution ($\Delta b = 10^{-4}$), the boundary cases of each capture band are circular.} 

\item{
In capture regions with consistent direction of rotation, eccentricity is a continuous function of $b$, as is seen in the region $b\in [1.904,1.914]$. The continuity of eccentricity with respect to impact parameter in these regions distinguishes our results from the friction-less case, investigated in \citet{boekholt_jacobi_2022}, in which the behavior of binary interactions were fractal-like with respect to impact parameter. In fact, by
examining encounters separated by sufficiently small $\Delta b$, we start finding a nearly unchanging eccentricity (shown in the upper right panel of \autoref{fig:no_fractal}). Plotting two simulations run with a small difference in initial $b$ of $2 \times 10^{-3}$, in the lower panels of \autoref{fig:no_fractal}, we find that they are very similar, with a visible deviation near the first pericenter. These results suggest that eccentricity and orbital paths are smooth functions of $b$ when DF is included.
}

\item{
Eccentricity, though itself continuous in regions with the same sense of rotation, has discontinuous jumps. This is well illustrated in \autoref{fig:rot_capture_hist}. The gradual decrease in eccentricity from $b=1.345$ to $b=1.365$ is followed by a much more sudden rise at $b=1.37$. The eccentricities of prograde orbits oscillate in the region $b\in [1.6,1.8]$.}

\item{Extremely high-eccentricity binaries ($1-e\approx10^{-3}$) are associated with a switch in sense of rotation. Each switch from prograde to retrograde or vice-versa is marked by very eccentric binaries rotating in either direction (see \autoref{fig:rot_capture_hist}). In the case of our fiducial suite of simulations we note that the switch from one rotation to another is more aptly called a ``transition'' in that it contains binaries that switch directions of rotation and have discontinuous jumps in eccentricity (see transition near $b=1.58$). At such high eccentricities, small differences in tidal force or DF will significantly affect the binary's orbit and possibly cause the discontinuity of eccentricity and direction of rotation we see in the aforementioned ``transition region''.}

\item{Captured orbits with undetermined features tend to be found in transition regions, near orbits of high eccentricity. Blue orbits are orbits that our integrator was not able to integrate to the simulation stopping condition. We determined that these blue orbits can be considered captured by our energy-based criterion (see \autoref{ssec:timetracing}). We have found that generally these blue orbits displayed high eccentricity and very close approaches ($<10^{-6}R_{\rm{Hill}}$) before the integration failed due to reaching exceedingly small small time-steps. The inability to resolve blue orbits does not affect our main conclusions.}

\end{enumerate}

\section{Dependence on system parameters}
\label{sec:astrophysical_reality}

In the previous section, we focused on the outcome (that is, capture {\it vs.} no-capture) of a close fly-by as a function of the impact parameter.  In order to assess the global prevalence of gas-induced binary capture, in this section we discuss how capture depends on the other system parameters, i.e. the BH masses, disc temperature and density, and the overall strength of dynamical friction.

\subsection{Scaling the fiducial model}\label{ssec:degenerate_orbits}

First, we note that our fiducial model can be directly scaled to be applicable to other parameter combinations.  We discuss this scaling in this subsection.

In the absence of any dynamical friction, the equation of motion in the Hill frame can be made dimensionless by measuring the separation between the two small bodies in units of their Hill radius $(m_2/M_0)^{1/3}a$, and time in units of the orbital time $n^{-1}=(GM_0/{a^{3}})^{-1/2}$ around the SMBH, where $a \equiv r_{0,1}+(b/2)\sqrt[3]{3}R_{\rm{Hill}}$. For example, assuming $m_1, m_2 \ll M_0$,
in the co-rotating coordinate system, the $x$-component of Hill's equations is
\begin{equation}
\Ddot{x} - 2n\dot{y} -3n^2 x = -\frac{\partial\phi}{\partial x},
\label{eqn:eom}
\end{equation}
where $x=x_1-x_2$ is the component of the separation $\vec{r}=\vec{r}_{0,2}-\vec{r}_{0,1}$ of the two small bodies along the $x$ axis pointing radially away from the SMBH, and $\phi=-G(m_1+m_2)/r$ is their interaction potential.   Introducing dimensionless distances $(r^\prime,x^\prime,y^\prime)\equiv (r/R_{\rm{Hill}},x/R_{\rm{Hill}},y/R_{\rm{Hill}})$ and time $t^\prime\equiv tn$ leads to 
\begin{equation}
(R_{\rm{Hill}}n^2)(\Ddot{x^\prime} - 2\dot{y^\prime} -3 x^\prime) = -\left(\frac{Gm_2}{R_{\rm Hill}^2}\right) \left(\frac{\partial\phi^\prime}{\partial x}\right),
\label{eqn:eomprime1}
\end{equation}
but since the dimensionful first terms on the left and right-hand sides are equal, they cancel, yielding the dimensionless equation
\begin{equation}
\Ddot{x^\prime} - 2\dot{y^\prime} -3 x^\prime = -\frac{\partial\phi^\prime}{\partial x},
\label{eqn:eomprime}
\end{equation}
where $\phi^\prime \equiv 1/r^\prime$.

Adding dynamical friction to the equations of motion breaks their scale-invariance in general.  However, in the subsonic regime ($v_{m}/c_{s} \ll 1$), the pre-factor in \autoref{eqn:ostriker} becomes
\begin{equation}\label{eqn:vel_relation}
    \frac{f(\frac{v_{m}}{c_{s}})}{{(\frac{v_{m}}{c_{s}})}^{2}}=\frac{v_{m}}{3c_{s}}.
\end{equation}
Substituting \autoref{eqn:vel_relation} into \autoref{eqn:ostriker} we can write the acceleration of $\rm{sBH}_{2}$ due to DF as
\begin{equation}\label{eqn:df_vrel}
    \vec{a}_{\rm DF} = \frac{ 4\pi^{2}G^{2} m_{2} \rho \vec{v}_{2} }{ 3 {c_{s}}^{3}}.
\end{equation}
Adding this term to the equation of motion, and using the
dimensionless distances and times results in a new term on the left-hand side of \autoref{eqn:eom}, $(4 \pi^2 G^2 m_2 \rho / 3 c_s^3)\dot{x} = (4 \pi^2 G^2 m_2 \rho / 3 c_s^3)(R_{\rm{Hill}}n)\dot{x^\prime}$. As long as the constant in front of $\dot{x^\prime}$ equals the previous dimensionful constants $R_{\rm{Hill}}n^2=Gm_2/R_{\rm Hill}^2$, these constants still all cancel, and the dimensionless form (\autoref{eqn:eomprime}) is preserved, i.e.
\begin{equation}
\Ddot{x^\prime} - 2\dot{y^\prime} - \dot{x^\prime}-3 x^\prime = -\frac{\partial\phi^\prime}{\partial x},
\label{eqn:eomprimedf}
\end{equation}
as long as
\begin{equation}\label{eqn:fric_ratio_full}
    \frac{ (2\pi)^{3} G^{3/2} m_{2}\rho a^{3/2}}{3 c_{s}^{3} M_{0}^{1/2}} = 1.
\end{equation}

This last equation is equivalent to intuitive condition that the normalization coefficient of the DF force (i.e. the quantity multiplying the velocity $v_m$) is a fixed fraction of the force between the two sBHs,
\begin{equation}\label{eqn:proportion_ratio}
  \frac{a_{\rm{DF}}}{a_{1,2}} \propto \frac{m_{2}\rho a^{3/2}}{ c_{s}^{3} M_{0}^{1/2}} = {\rm constant}. 
\end{equation}
As long as this condition is satisfied, 
and the initial (dimensionless) impact parameters $b$ are identical, the orbits in the Hill frame, measured in Hill units, will be indistinguishable from those in our fiducial model. This allows us to generalize our fiducial results for a wide range of systems. Systems with the same $b$ and same value for the ratio \autoref{eqn:proportion_ratio} will have the same capture result, direction of rotation, and eccentricity evolution, etc.

We have numerically verified this scaling by running a suite of simulations that individually varied each parameter in \autoref{eqn:proportion_ratio} by $10^{0.1}$ along a range of values that were $10^{3}$ greater and less than their fiducial value (see \autoref{tab:fiducial_params}), effectively creating a lattice of simulations along 5 axes. We then iterated though all possible permutations of this lattice's 2d projections and found that the points (in the 2-d plane) which fell along the curves dictated by the relations in \autoref{eqn:proportion_ratio} (i.e $m_2 \propto {c_s}^{3}$ in the $m_2,c_s$ plane, $\rho \propto {M_0}^{1/2}$ in the $\rho, M_0$ plane, $\rho \propto {m_2}^{-1}$ in the $\rho, m_2$ plane, etc.) yielded identical orbits\footnote{
The orbits are identical (``degenerate'') only for initial $\Delta \phi \gtrsim 100R_{\rm{Hill}}$. The large azimuthal separation ensures that the scaling across simulations is not broken by the slight gravitational forces of the companion at the starting positions. Since $\Delta \phi$ only affects the scaling between simulations, our discussion of the properties of orbits with $\Delta \phi = 10R_{\rm{Hill}}$ remains valid.} in Hill space.  Deviations from this scaling begin to be discernible only when the sBH's orbital Mach numbers, relative to the background disc, reach values of $v_2/c_s\gtrsim 0.3$.

\subsection{Dependence of capture on the strength of DF}\label{ssec:cap_tol}

In the previous section, we showed that orbits with the same ratio of the DF friction force normalization to the mutual gravitational force (\autoref{eqn:proportion_ratio}) follow the same path in the dimensionless Hill frame, when initialised with the same impact parameter $b$. Conversely, when either this ratio (which we hereafter refer to as the friction force ratio, or FFR) or the impact parameter $b$ is changed, the orbital path changes.   

\begin{figure}
    \centering
    \includegraphics[width=1\columnwidth]{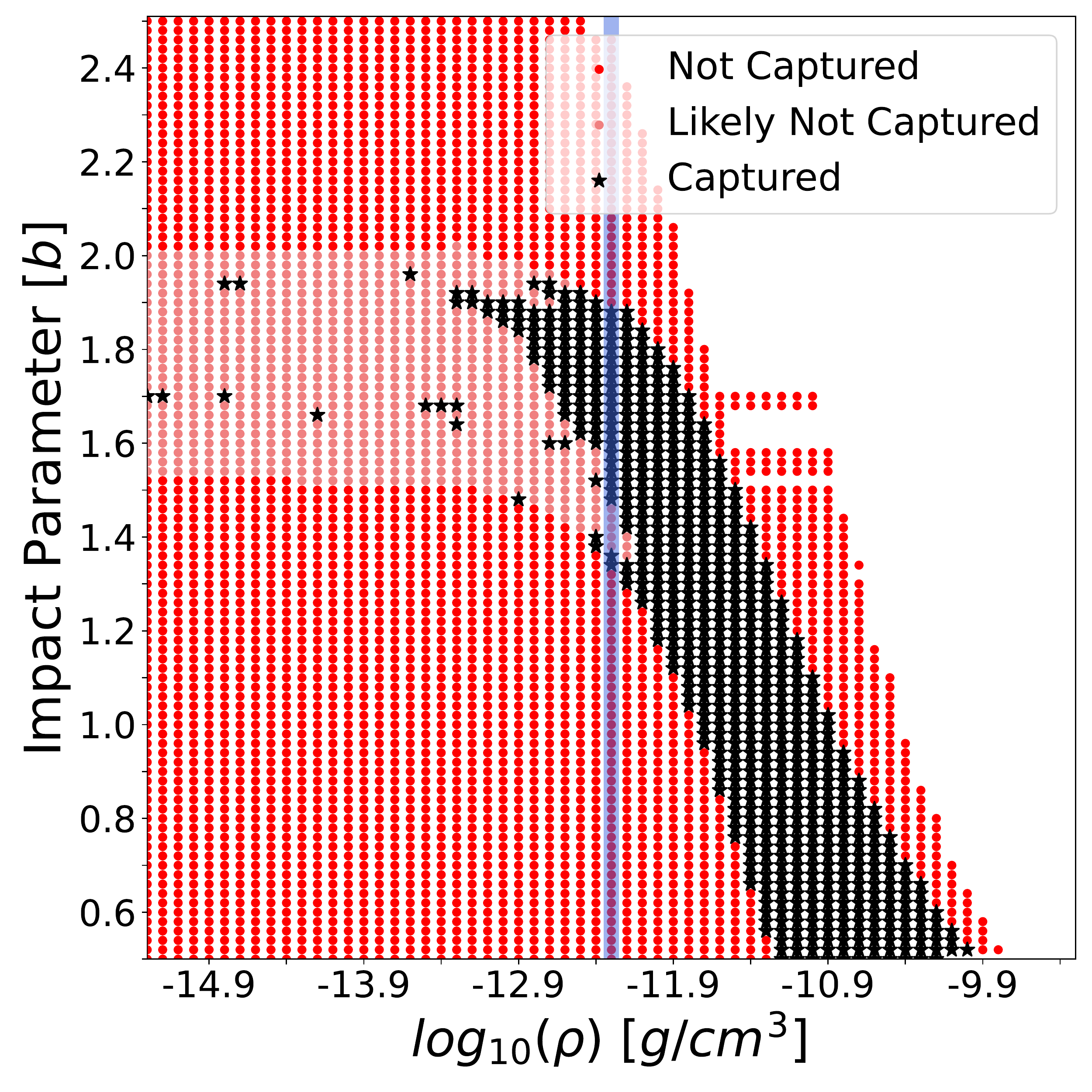}
    \caption{Capture occurrence for a larger suite of lower-precision simulations. Each colored point on the graph represents an executed simulation. The red dots indicate runs that did not result in a capture, black stars indicate runs that did, while pink dots indicate runs that likely did not result in capture. The simulations were run using our fiducial system parameters and only varied $b$ and $\rho$. The vertical line shaded by blue at $\rho=10^{-12.3} \rm{g}\,\rm{cm}^{-3}$ marks our fiducial run.}
    \label{fig:rholines}
\end{figure}

In \autoref{ssec:density_vary} we noted that varying the disc density has a distinct effect on capture occurrence. Namely, at fixed $b$, if the density is either too high or too low, the sBHs are not captured. However, we now recognize that the density is just a proxy for the FFR -- i.e. it is the combination of BH ($M_0, m_2, a$) and disc ($\rho, c_s$) parameters in \autoref{eqn:proportion_ratio} that determines the orbit, rather than just the density $\rho$.  By running a suite of simulations keeping all parameters at their fiducial values, except varying both $b$ and $\rho$, we are therefore able to constrain capture occurrence in the two-dimensional $b$-FFR space. 

We performed a suite of simulations with $\rho\in[10^{-9.3},10^{-15.3}]\,{\rm g\, cm^{-3}}$ sampled with a step size $\Delta \rho=10^{0.1}\,{\rm g\, cm^{-3}}$ and $b\in[0.5,2.5]$ at each density/FFR  sampled with step size $\Delta b=0.02$. Determining whether the sBHs are captured for each simulation in this suite determines capture occurrence for any ($M_0, m_2, a, \rho, c_s$) parameter combination as long as sBHs move subsonically with respect to the AGN disc gas, which we typically find to be the case (see \autoref{ssec:degenerate_orbits}).

Adaptive integration of systems with higher FFR has a higher computational cost per orbit to achieve the same accuracy. To avoid lengthy computation times, we increase the precision parameter from \verb_Epsilon_ $=10^{-8}$ in our fiducial runs to  \verb_Epsilon_ $=10^{-4}$ in this suite. This, however, produces many orbits whose fate, based on our graphical criterion, remains ambiguous (\autoref{ssec:eccrot_rh}). We instead determine capture by the simpler orbital energy criterion (\autoref{ssec:timetracing}). Naively, all simulations with a minimum energy below zero are captured, while those with energies remaining positive throughout the orbit are not. However, as we noted earlier, the changing radial separation of the binary creates small oscillations in its energy. Thus, we impose a more stringent criterion, and identify runs in which the binary energy dips below a negative threshold\footnote{The value of this threshold was determined by comparing the capture occurrence results using the energy criterion with those produced by our graphical capture criterion in our fiducial runs and choosing the threshold value for which the two match.} of $E\leq -2.11 \times 10^{-2}{\rm M}_{\astrosun}{\rm AU\,Yr^{-2}}$.  
These runs are declared as {\em captured}; those with positive minimum binary energy as {\em not captured}, and those with minimum energy values in-between as {\em likely not captured}. The results of this analysis are shown in \autoref{fig:rholines}.

\autoref{fig:rholines} reveals that the gas-capture mechanism is not restricted to our fiducial $\rho$/FFR.We see that the vertical $b$-interval for capture that we identified earlier at fixed $\rho$ is just a 1D slice of a two-dimensional capture band in the ($b,\rho$) plane.  At fixed $b$, this band has a horizontal width of approximately an order of magnitude, i.e. $\Delta \rm{FFR}=\Delta \rho\approx10^{1.2}$ (or equivalently $\Delta M_{0}\approx10^{2.4}$,  $\Delta c_{s}\approx10^{0.4}$, etc). Gas capture apparently can create bound binaries for a range of impact parameters of order the Hill radius, and for an order-of-magnitude range of the DF friction force.

\autoref{fig:rholines} also directly confirms the inverse monotonic trend between $b$ and $\rho$ we noted in \autoref{ssec:density_vary}.  However, we can now generalize this conclusion as a trend between $b$ and
the FFR combination of ($M_0, m_2, a, \rho, c_s$). First, the figure confirms that the $b$ capture range shifts to lower values with increasing FFR, due to the changing balance between mutual gravity and relative friction between the sBHs. Additionally, we note that at $\rho=10^{-11.4}{\rm g\,cm^{-3}}$ the width of the band is $\Delta b = 0.76$ while at $\rho=10^{-12.1}{\rm g\,cm^{-3}}$, it is somewhat narrower, $\Delta b = 0.58$. Apparently, when DF is stronger, a larger range in $b$ can be captured. Similarly, the width of the FFR capture range increases at smaller impact parameters $b$. These trends can be understood qualitatively by noting that the same fractional change in friction will have a smaller effect on a system with a larger mutual gravity.

We also observe that \autoref{fig:rholines} exhibits both continuous and discrete capture bands in $b$, depending on the value of $\rho$. We see discrete capture bands for $\rho \lesssim 10^{-12.1} {\rm g\,cm^{-3}}$, while the capture bands become continuous for $\rho \gtrsim 10^{-12.1}{\rm g\,cm^{-3}}$. This trend, too, can be qualitatively explained by the interplay of relative friction levels and mutual gravity. It is conceivable that at $\rho \approx10^{-12.1} {\rm g\,cm^{-3}}$ the relative magnitude of friction is on the verge of being large enough that variations in $b$ do not change the orbital paths enough that capture becomes discontinuous. 

Our results build on the work of \citet{boekholt_jacobi_2022} in which the longevity of binaries captured absent DF was found to have a fractal-like structure across impact parameter ($b$). The addition of DF extends the fractal `islands' of capture, creating contiguous bands of capture across the sampled $b$ (\autoref{fig:rholines}). By reducing the DF coefficient to sufficiently low values, capture regions shrink and separate, suggesting continuity between the low and no-friction cases.

\section{Implications}\label{sec:implications}

In this section we discuss the implications of our results for AGN disc models, and for previous and future work on binary formation through gaseous dynamical friction in AGN discs. In \autoref{ssec:capturediscs} we give explicit estimates for where captures would occur in a physical AGN disc model. In \autoref{ssec:sams} we compare our results to a simple previously adopted semi-analytic recipe for binary formation. Finally, in \autoref{ssec:compare}, we discuss similarities and differences between our study and related previous works.

\subsection{Captures in physical AGN disc models}\label{ssec:capturediscs}

\begin{figure}
    \centering
    \includegraphics[width=1\columnwidth]{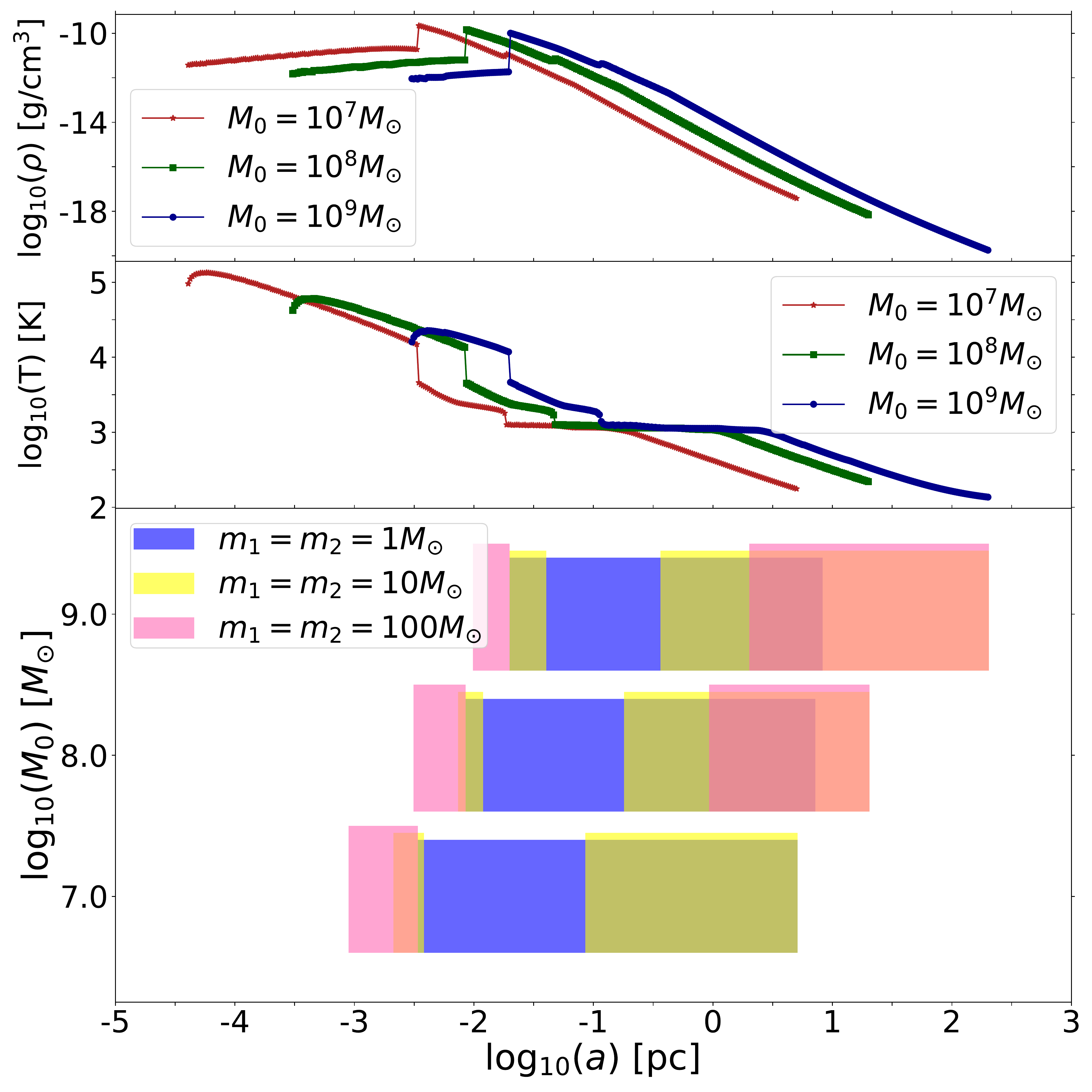}
    \caption{Expected locations of captures in an AGN disc. The two upper panels show the radial profiles of density
    ($\rho$) and mid-plane temperature ($T$) for AGN disc models with SMBH masses of $10^{7,8,9}M_{\astrosun}$. Models are built following \protect \citet{thompson_radiation_2005}, with gas supply rates to the outer edge of the disc equal to $320$, $32$, $1.5 \text{ M}_\odot/$year and accretion rate at the inner edge of the disc equal to $0.19$, $0.98$, $4.8 \dot{\text{ M}}_{\text{Edd}}$ from the most massive to the least massive SMBH respectively. The lower panel represents the range of positions in the disc $a$ where capture occurs. Each horizontal bar corresponds to a different choice of SMBH mass $M_0$ as labeled on the $y$ axis, and for different sBH masses $m_2$ as labeled in the legend.}
    \label{fig:real_agn_caps}
\end{figure}

\begin{figure}
    \centering
    \includegraphics[width=1\columnwidth]{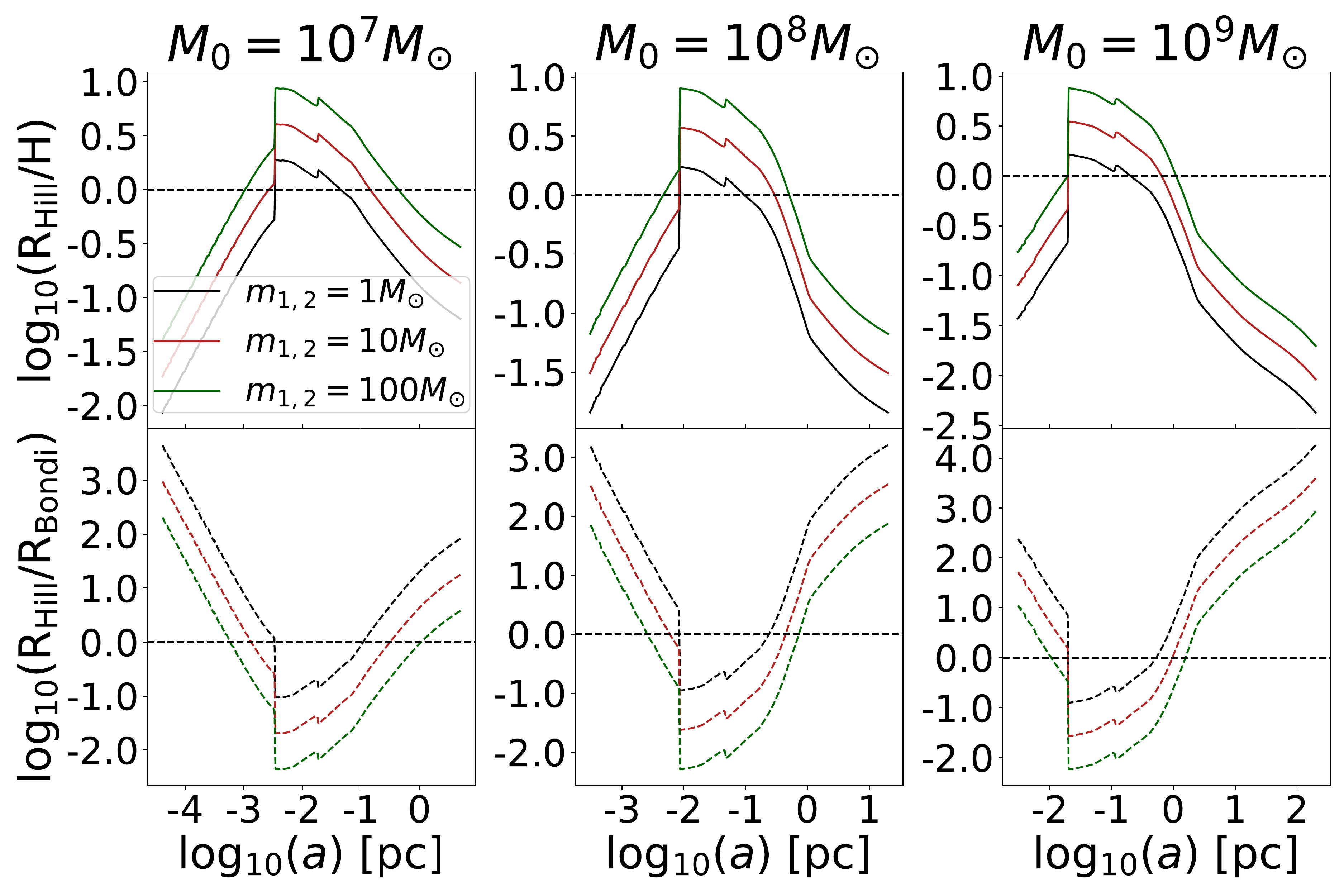}
    \caption{Radial regions in the AGN disc where our DF model is unphysical. The upper row shows the ratio $R_{\rm{Hill}}/H$ as a function of the binary's radial position in the AGN disc, for each (SMBH + sBH$_1$ + sBH$_2$) system depicted in \autoref{fig:real_agn_caps}. The bottom row shows instead the ratio $R_{\rm{Hill}}/R_{\rm{Bondi}}$. Our model is unphysical when $\rm{log}_{10}(\rm{R}_{\rm{Hill}}/\rm{H})$ is above the $y=0$ line or when $\rm{log}_{10}(\rm{R}_{\rm{Hill}}/\rm{R}_{\rm{Bondi}})$ is below it.}
\label{fig:geometric_discs}
\end{figure}

Given our constraints on capture in $b$-$\rho$ space, we note that the range $\rho \in [10^{-12.8}, 10^{-10.4}]$ g cm$^{-3}$ generally results in capture for some range of $b$ with $\Delta b \le2$. Any (SMBH + sBH$_1$ + sBH$_2$) triple system with the range of FFR values that corresponds to this density range in our fiducial model will therefore result in capture, as well. Physical models of geometrically thin AGN discs yield the radial profiles $\rho=\rho(a)$ and $c_{s}=c_{s}(a)$ for a given $M_0$ (once the overall accretion rate and viscosity in the disc is chosen).  Using a physical model, we can then calculate the FFR, for any given choices of $M_0$ and of $m_1=m_2$, as a function of the location $a$ in the disc.

In \autoref{fig:real_agn_caps}, we use this approach to estimate where in AGN discs binaries can form via gas capture.  In particular, we adopt the AGN disc models from \citet{thompson_radiation_2005}, parameterized by the SMBH mass ($10^{7,8,9}M_{\astrosun}$), the outer-most disc radius ($5$, $20$, $200$~pc), and the gas supply rate at the outer edge of the disc ($1.5$, $32$, $320 \text{ M}_{\odot}/\text{ year}$). We assume angular momentum transfer proceeds via global torques, such that the radial velocity remains a constant fraction of the sound speed or $v_{\rm r} = c_{\rm s} m$ where we set $m = 0.2$. Defining the Eddington accretion rate as $\dot{M}_{\text{Edd}} = 10\times L_{\text{Edd}}/c^2$, the resulting SMBH mass accretion rates are $4.8$, $0.98$, $0.19 \dot{M}_{\text{Edd}}$. The density and mid-plane temperature profiles for these models are shown in the upper panel of \autoref{fig:real_agn_caps}. In the lower panel, we plot the ranges of $a$ corresponding to the aforementioned FFR range of capture, for three different (equal) sBH masses of $1$, $10$, and $100$ M$_{\astrosun}$, and for the three different SMBH masses of $10^{7,8,9}$ M$_{\astrosun}$, in this AGN disc model.

In addition to confirming that captures can occur in physical AGN disc models, \autoref{fig:real_agn_caps} illustrates some interesting trends. First, captures can occur throughout a wide range of locations $a \in [10^{-3}, 10^{2.5}]$~pc.  Second, the annuli of capture for different $m_2$ values overlap. This makes sense as AGN profiles, largely, only display gradual changes with respect to position in the disc, thus it is expected that we will see gradual transitions between capture occurrence for different mass sBHs. Thirdly, more massive sBHs experience capture towards either end of the disc as compared to their lower-mass counterparts. Interestingly, there are two distinct annuli where captures happen for some sBH/SMBH mass combinations; this is a result of the particular temperature and density profiles, caused by sharp opacity changes in our chosen AGN disc model.

The calculations to produce \autoref{fig:real_agn_caps} are meant to be illustrative, in a specific AGN disc model, with a single choice of accretion rate and viscosity. The conditions for capture identified in \autoref{fig:rholines} and \autoref{eqn:proportion_ratio} are more general, and could be easily applied to our chosen AGN disc model with different assumed accretion rates and viscosity, and also to other AGN disc models which predict different $\rho$ and $c_s$ profiles.

Unlike in our fiducial model, there are regions in the aforementioned AGN disc models (see \autoref{fig:real_agn_caps}) where our dynamical friction prescription (see \autoref{ssec:dyfric}) is no longer physical. Namely, when $R_{\rm{Hill}} > H$ we may be in a gap opening regime \citep{Crida_2006, Dempsey_2022}, when $R_{\rm{Hill}}$ < $R_{\rm{Bondi}}$, the accretion onto the black hole is tidally limited \citep{Dittmann_2021}, and when $\rm{min}\{R_{\rm{Hill}},R_{\rm{Bondi}}\} > H$ the cross section of the wake formed by the motion of the sBHs in the disc will be reduced and the full Ostriker drag will not be realised.

In \autoref{fig:geometric_discs} we compare the values of $R_{\rm{Bondi}}$, $R_{\rm{Hill}}$, and $H$ as functions of $a$ for various SMBH and sBH masses. First, we note that all the systems in \autoref{fig:real_agn_caps} have regions where our DF is unphysical. Second, we find that the radial regions where R$_{\rm{Hill}}$ > H and where R$_{\rm{Hill}}$ < R$_{\rm{Bondi}}$ coincide nearly exactly, leaving a single region in each disc where we are over-estimating the frictional effect of the wake. Third, we find that the degree to which our prescription of DF is unphysical is largely independent of the mass of the SMBH, and depends only on the binary's mass. Notably, the unphysical range of radii is narrower for lower sBH masses. Finally, we note that the regions where our DF force is unphysically large nearly line up with the regions in which the corresponding systems do not capture sBHs in \autoref{fig:real_agn_caps} -- namely the inner region of the disc where $a \in [\approx 10^{-2.5}, \approx 10^{0}]$~pc. This is of particular note since if we were to weaken the magnitude of DF to reflect the smaller cross section of the wake in these regions, we may lower the high FFR values in these regions to those where capture can occur. Thus, a more nuanced account of the geometry of the disc and the Bondi radius may predict more captures throughout the inner region of the disc--rendering the annuli of capture in \autoref{fig:real_agn_caps} to be largely continuous with few gaps.

\subsection{Implications for semi-analytic models and recipes}\label{ssec:sams}

\begin{figure}
    \centering
    \includegraphics[width=1\columnwidth]{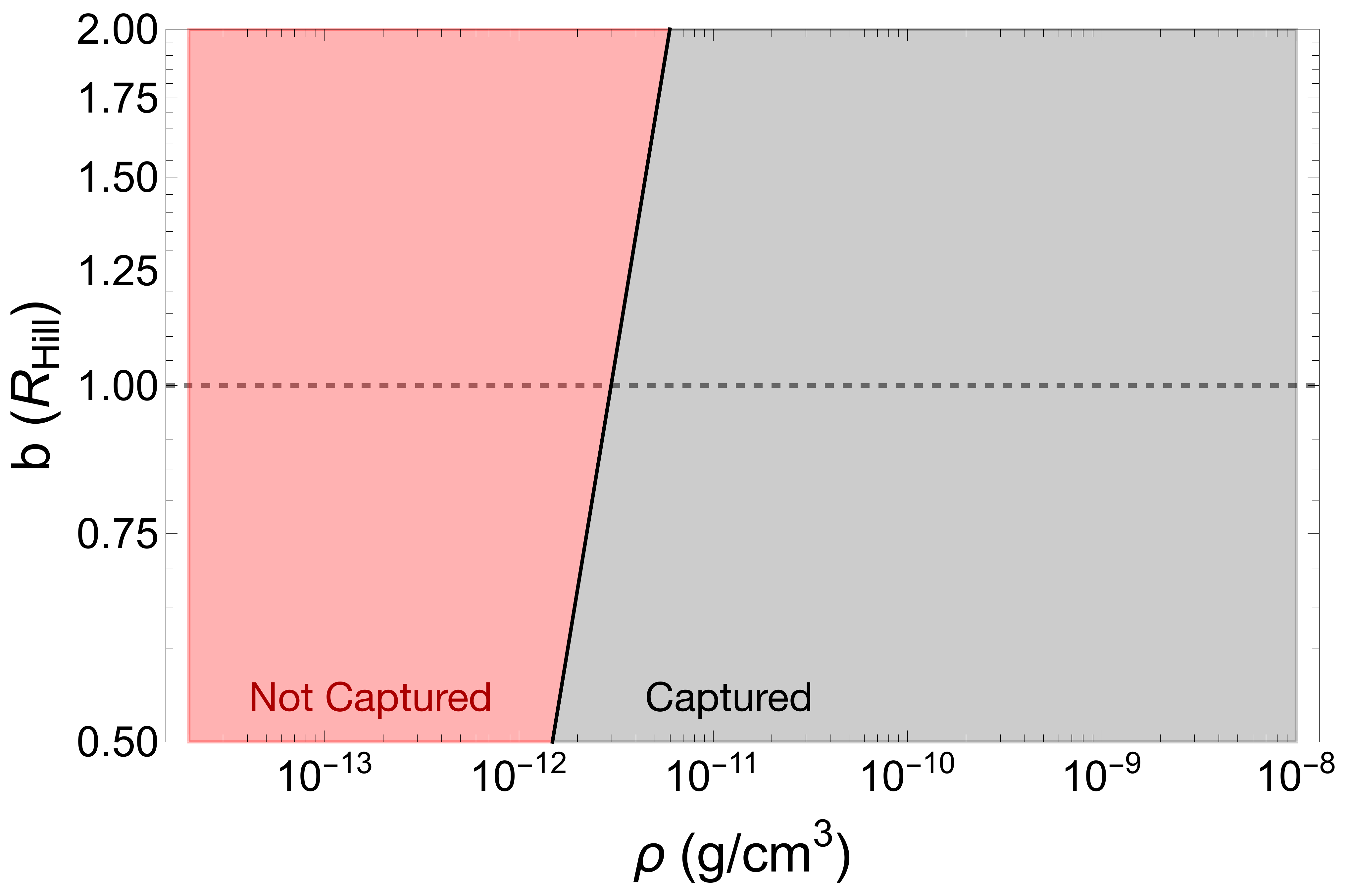}
    \caption{Capture boundary calculated according to the analytical prescription suggested by \protect\citet{tagawa_formation_2020} and \protect\citet{goldreich_formation_2002} i.e. $t_{\text{pass}}/t_{\text{DF}}\gtrapprox 1$. Regions in which capture is expected are filled in gray, and regions without capture are filled in red. The dashed line marks $R_{\rm{Hill}}$, the capture limit set by \protect\citet{tagawa_formation_2020}.  We note that the gray capture region is effectively truncated to exclude large $b$ by Equation 40 in \protect\citet{tagawa_formation_2020}. The implied capture regions differ markedly from those obtained in our orbital calculations (cf.~\autoref{fig:rholines}). See \autoref{ssec:sams} for discussion.}
    \label{fig:tagawa_capture}
\end{figure}

When friction is large relative to sBH mutual gravity (the high FFR regime), sBHs cannot deviate significantly from their Keplerian trajectories, and are unable to form bound binaries. This result suggests a more conservative estimate of capture occurrence than the simplified scaling derived in \citet{goldreich_formation_2002} and \citet{tagawa_formation_2020}. In the formulation laid out by \citet{tagawa_formation_2020}, i.e. in their equations (62 - 64), the probability of binary capture may be approximated roughly by the ratio of the $R_{\text{Hill}}$ crossing time ($t_{\text{pass}} = R_{\text{Hill}}/v_\text{rel}$) to the timescale of damping of the relative velocities between the sBHs ($t_{\text{GDF}} = v_{\text{rel}}/a_{\text{DF}}$). From this prescription we may infer that the capture boundary is set roughly by $t_{\text{GDF}}/t_{\text{pass}} \sim 1$. Re-writing this ratio in terms of $v_{\text{rel}}$ and familiar simulation parameters and constants we calculate 
\begin{equation}
    v_{\text{rel}} \sim \frac{4 \pi}{3^{2/3}} \frac{G^2 a \rho m_1^{2/3}}{c_{\text{s}}^3 M^{1/3}}\,.
\label{eqn:Hiromichi_capture}
\end{equation}
In all of our simulations, $v_{\text{rel}}$ is initially set to the Keplerian shear velocity, and may therefore be understood as a proxy for $b$. In \autoref{fig:tagawa_capture} we substitute the Keplerian velocities of the sBHs for $v_{\text{rel}}$ in \autoref{eqn:Hiromichi_capture}, and plot the boundary of capture in terms of $b$ and $\rho$. As in \autoref{fig:rholines}, the gray shaded region indicates capture. The horizontal dashed line at $b = 1$ in \autoref{eqn:Hiromichi_capture} indicates where \citet{tagawa_formation_2020} capped binary formation. The location of the capture boundary near the fiducial density and impact parameter $b\sim 1$ is in rough agreement with our results. Note, however, that the trends in $b$ {\it vs.} $\rho$ are the opposite, and \autoref{fig:tagawa_capture} suggests that for any given $b$, there is a minimum $\rho$ above which all encounters result in capture, thereby overestimating capture in the high $\rho$ regime. Ultimately, this overestimation may not have a significant effect on the binary formation rates found by \citet{tagawa_formation_2020}, whose disks are modeled according to \citet{thompson_radiation_2005} and do not exceed densities of $\sim 10^{-10}\text{ g/cm}^3$. Still, we recommend that future semi-analytic models of AGN disc-embedded sBH binary populations adopt the markedly different capture criteria implied by \autoref{fig:rholines} and \autoref{eqn:proportion_ratio}.

\subsection{Comparison to other works}\label{ssec:compare}

During the final stages of completing this manuscript, we became aware of a related study by \citet{LiDempseyLi+2022}, addressing binary capture using two-dimensional hydrodynamic simulations.  They found that sBHs with $\Delta b<2$ are captured and form bound binaries.  In analysing the close encounters, they find that the strong interaction and collision between the two sBHs' minidiscs has the largest effect in binding the binaries.  This contrasts somewhat with our demonstration that bound binaries form already on the approach to the close interaction. It is unclear whether this apparent difference arises from the difference between the full hydrodynamics and our simplified treatment of dynamical friction, or whether the difference is in post-simulation analyses. Furthermore, \citet{LiDempseyLi+2022} find that once formed, all of their captured binaries are on retrograde and eccentric orbits, whereas we find that prograde and circular binaries can also form.  While the hydrodynamic study is a more realistic description of the sBHs' encounter and their interaction with the disc gas, we also note that \citet{LiDempseyLi+2022} examined only a single impact parameter and did not fully explore the phase space of \autoref{fig:rholines}. Therefore, further simulations are required to assess whether prograde and circular binaries arise in AGN discs. \citet{LiDempseyLi+2022} summarise their results in terms of a capture criterion, defined in the two-dimensional plane of the disc/sBH mass ratio {\it vs.} the binary energy at a fixed separation of $b=0.3$. This criterion remains distinct from ours outlined in \autoref{ssec:cap_tol} and comparisons between the two are inconclusive. On the other hand, \citet{LiDempseyLi+2022} do not explore the parameter space of even higher disc densities, where we find dynamical friction not to lead to captures. Overall, it is reassuring that both treatments find that captures can happen and do not appear to require a special fine-tuning of parameters.

\citet{li_long-term_2022} also study gaseous dynamical friction capture of sBHs in AGN discs using an N-body numerical code. They adopt a simple analytic expression for DF that describes the drag effects via a characteristic timescale $\tau_{\rm{DF}}$,
\begin{equation}
    \vec{a}_{\rm{DF}}=\vec{v}_{m}/\tau_{\rm{DF}}\,,
    \label{eqn:li_friction}
\end{equation}
or equation 22 in \citet{li_long-term_2022}. Noting that $a_{\rm{DF}} \propto v_{\rm{m}}$ in \autoref{eqn:li_friction}, we may substitute our subsonic prescription, \autoref{eqn:df_vrel}, for $a_{\rm{DF}}$ and solve for $\tau_{\rm{DF}}$:
\begin{equation}
    \tau_{\rm{DF}}=\frac{ 3 {c_{\rm s}}^{3}}{ 4\pi^{2}G^{2} m_{2} \rho}\,.
\end{equation}
In our fiducial set-up, the approximate range of capture lies between $\rho \in \left[-12.8, -10.4\right]$ with a corresponding drag timescale between  $\tau_{\rm DF}\in \left[308, 1.23\right]$ years.  We can thus compare our results to those of \citet{li_long-term_2022}, who assume $\tau_{\rm DF} = 10^5\times \tau_{\rm sBH_{1}}$ and $10^6\times \tau_{\rm sBH_{1}}$ or $3 \times 10^7$ and $3 \times 10^8$ years for $M = 10^8 \text{ M}_{\odot}$ and $a = 0.1$ pc. The $\tau_{\rm DF}$ for which our simulations result in bound binaries is orders of magnitude smaller than the timescales employed by \citet{li_long-term_2022}, who find that close encounters do not result in long-lived binaries. These results are not in conflict with our simulations, because they probe a region of parameter space in which we expect the dynamics to be well described by the frictionless case.

\section{Summary and Conclusions}\label{sec:conclusions}

In this paper, we have examined the capture of sBHs in gaseous accretion discs under the influence of DF, using direct orbital integrations, with DF modeled through a commonly used fitting formula from \citet{ostriker_dynamical_1999}. The summary of our main results are as follows:

\begin{enumerate}
\item{A pair of single sBHs embedded in an AGN disc, approaching each other in the impact parameter range $b\in [0,2]$ in Hill radius units can form a bound binary when gaseous dynamical friction is accounted for.  Significantly more binaries can be captured with DF than in the absence of DF \citep{boekholt_jacobi_2022, li_dynamical_instability_2022}}.

\item{The binary already becomes bound due to the energy that was dissipated by DF on the approach just before the first close interaction of the pair of sBHs.}

\item{Capture occurs only when mutual gravity and DF do not overwhelm the effects of the other. Very high and low levels of DF both hinder capture, suggesting a more conservative estimate of capture occurrence than suggested by semi-analytic models \citep{goldreich_formation_2002, tagawa_formation_2020}}.   When all other parameters are held fixed, this means captures occur for a finite range of background AGN disc densities.

\item{DF creates regions (``bands'') of capture in the two-dimensional plane of impact parameter ($b$)  {\it vs.} a quantity we defined as
the friction force ratio (FFR).  FFR represents a
combination of ($M_0, m_2, a, \rho, c_s$) that keep the ratio of the DF force coefficient to the mutual gravitational force between the sBHs constant (\autoref{eqn:proportion_ratio}).  In this  $b$-FFR plane, there is a contiguous region of capture (\autoref{fig:rholines}). At low FFR values capture only occurs for narrow discrete regions in $b$, suggesting a smooth transition between the low-friction and frictionless simulations \citep{boekholt_jacobi_2022, li_dynamical_instability_2022}}.

\item{Bound binaries exhibit a wide range of eccentricities and sense of rotation. Prograde and retrograde orbits are produced approximately equally frequently, suggesting that DF forms binaries from near Hill velocity approaches \citep{schlichting_ratio_2008}. Prograde binaries are preferentially very eccentric, while retrograde binaries have a broad range of near-circular to eccentric orbits.}
    
\item{Orbits in simulations that have the same FFR = $(m_1\rho a^{3/2})/(c_s^3 M_{0}^{1/2})$ (\autoref{eqn:proportion_ratio}) and the same $b$ are identical in dimensionless units, with time measured in orbital time and  distance in Hill radii. This holds as long as the sBHs have subsonic velocities relative to the AGN disc gas, which is typically the case until the very strong close interacton of the two sBHs.  This scaling allows our capture boundaries to be scaled to different combinations of the AGN and sBH parameters.}

\item{In particular, gas captures occur between $10^{-3}$ - $10^{2.5}$ pc in illustrative physical AGN disc models, representing bright quasars with SMBHs with masses in the range of $10^{7-9}$ M$_{\astrosun}$, and with sBHs in the range of $1-100$ M$_{\astrosun}$ (\autoref{ssec:capturediscs}).}
\end{enumerate}

In future work we hope to be able to further address how sensitive our results are to the assumptions of the sBHs being co-planar and equal-mass. Additionally, since our work employs three-body simulations only, it is of particular interest to assess how our findings correspond to hydrodynamic simulations that cover a wider range in $b$-FFR space. Lastly, our analytic criteria for capture can be incorporated into improved global population models of the sBH-sBH merger population, to better assess the contribution of this pathway to the GW events observed by LIGO/Virgo.

\section*{Acknowledgments}

We thank Hiromichi Tagawa, Bence Kocsis, and Mordecai-Mark Mac Low for useful discussions. ZH acknowledges financial support from NASA grant 80NSSC22K082 and NSF grants AST-2006176 and AST-1715661. MEM achknowledges financial support from the GRFSD fellowship. We also acknowledge computing resources from Columbia University's Shared Research Computing Facility project, which is supported by NIH Research Facility Improvement Grant 1G20RR030893-01, and associated funds from the New York State Empire State Development, Division of Science Technology and Innovation (NYSTAR) Contract C090171, both awarded April 15, 2010.

\section*{Data Availability}
The data underlying this article will be shared on reasonable request to the corresponding author.

\bibliographystyle{mnras}
\bibliography{main}

% Don't change these lines
\bsp	% typesetting comment
\label{lastpage}

\end{document}